\documentclass[pdflatex]{sn-jnl}

\usepackage{graphicx}%
\usepackage{multirow}%
\usepackage{amsmath,amssymb,amsfonts}%
\usepackage{amsthm}%
\usepackage{mathrsfs}%
\usepackage[title]{appendix}%
\usepackage{xcolor}%
\usepackage{textcomp}%
\usepackage{manyfoot}%
\usepackage{booktabs}%
\usepackage{algorithm}%
\usepackage{algorithmicx}%
\usepackage{algpseudocode}%
\usepackage{listings}%

\begin{document}

\title[Article Title]{Terahertz emission from \texorpdfstring{$\alpha$}{}-W/CoFe epitaxial spintronic emitters}

\author*[1,2]{\fnm{Venkatesh} \sur{Mottamchetty}}\email{venkatesh769@gmail.com}
\author[1]{\fnm{Rimantas} \sur{Brucas}}\email{rimantas.brucas@angstrom.uu.se}
\author[2]{\fnm{Anna L.} \sur{Ravensburg}}\email{anna.ravensburg@gmail.com}
\author[3]{\fnm{Renan} \sur{Maciel}}\email{renanm@kth.se}
\author[4]{\fnm{Danny} \sur{Thonig}}\email{Danny.Thonig@oru.se}
\author[5]{\fnm{J\"urgen} \sur{Henk}}\email{juergen.henk@physik.uni-halle.de}
\author[6,7]{\fnm{Rahul} \sur{Gupta}}\email{rahulguptaphy@gmail.com}
\author[1]{\fnm{Arne} \sur{Roos}}\email{Arne.Roos@angstrom.uu.se}
\author[8]{\fnm{Cheuk} \sur{Wai Tai}}\email{cheuk-wai.tai@mmk.su.se}
\author[2]{\fnm{Vassilios} \sur{Kapaklis}}\email{vassilios.kapaklis@physics.uu.se}
\author*[1]{\fnm{Peter} \sur{Svedlindh}}\email{peter.svedlindh@angstrom.uu.se}

\affil[1]{\orgdiv{Department of Materials Science and Engineering}, \orgname{Uppsala University}, \orgaddress{\street{Box 35}, \city{Uppsala}, \postcode{75103}, \country{Sweden}}}

\affil[2]{\orgdiv{Department of Physics and Astronomy}, \orgname{Uppsala University}, \orgaddress{\street{Box 516}, \city{Uppsala}, \postcode{75120}, \country{Sweden}}}

\affil[3]{\orgdiv{Department of Applied Physics}, \orgname{Royal Institute of Technology (KTH)}, \city{Stockholm}, \country{Sweden}}

\affil[4]{\orgdiv{School of Science and Technology}, \orgname{\"Orebro University}, \city{\"Orebro}, \postcode{70182}, \country{Sweden}}

\affil[5]{\orgdiv{Institut f\"ur Physik}, \orgname{Martin Luther University Halle-Wittenberg}, \city{Halle}, \postcode{06099}, \country{Germany}}

\affil[6]{\orgdiv{Department of Physics}, \orgname{University of Gothenburg}, \city{Gothenburg}, \postcode{33620}, \country{Sweden}}

\affil[7]{\orgdiv{Department of Physics}, \orgname{Indian Institute of Technology (Indian School of Mines) Dhanbad}, \city{Dhanbad}, \postcode{826004}, \country{India}}

\affil[8]{\orgdiv{Department of Materials and Environmental Chemistry}, \orgname{Stockholm University}, \city{Stockholm}, \postcode{10691}, \country{Sweden}}

\abstract{Efficient terahertz (THz) generation in epitaxial $\alpha$-W/Co$_{60}$Fe$_{40}$ spintronic emitters is reported. Two types of epitaxial emitters have been investigated; $\alpha$ -W $(110)$ / Co $_{60}$Fe$_{40}(110)$ and $\alpha$ -W $(001)$ / Co $_{60}$ Fe $_{40}(001)$ deposited on single crystalline Al$_{2}$O$_{3}$($11\bar{2}0$) and MgO($001$) substrates, respectively. The generated THz radiation is about $10\%$ larger for $\alpha$-W$(110)$/ Co$_{60}$Fe$_{40}(110)$ grown on Al$_{2}$O$_{3}$($11\bar{2}0$), which is explained in part by that the $\alpha$-W$(110)$/Co$_{60}$Fe$_{40}(110)$ interface for this emitter is more transparent to the spin current due to the presence of \AA ngstr\" om-scale interface intermixing at the W/CoFe interface. Another contributing effect to the difference in THz emission is the presence of local strain in the $\alpha$-W$(001)$ layer grown on MgO($001$), which due to spin scattering and spin memory loss will reduce the spin current. Our results also reveal that the generation of THz radiation is larger when pumping with the laser light from the substrate side. Measurements of reflectance and transmittance spectra as well as Fresnel calculations for the studied film stacks show that this can be explained by a larger part of the laser light being absorbed in Co$_{60}$Fe$_{40}$ in this measurement configuration due to interference effects.}


\maketitle

\section{Introduction}

Terahertz (THz) radiation is typically assigned to the $0.3 - 30$ THz frequency range of the electromagnetic spectrum corresponding to a wavelength range of 1 mm to 10 $\mu$m. The terahertz frequency range lies between the microwave and near infrared regions of the electromagnetic spectrum and has traditionally been called the terahertz gap because of the difficulties in generating THz radiation. This has limited its use despite the large number of possible applications \cite{valuvsis2021roadmap,federici2010review,Leitenstorfer_2023}. Techniques used for the generation of THz radiation include photoconductive switching, optical rectification, transient photocurrent in air plasma, and difference frequency generation \cite{smith1988subpicosecond,VENKATESH2014596,lepeshov2017enhancement,yang2007large,Kim2021,kim2007terahertz,yan2017high,lu2017modulation,fulop2020laser}. A more recent technique builds on the spin degree of freedom in magnetic heterostructures \cite{seifert2016efficient,gupta2021co2feal,PapaioannouBeigang,Kefayati}.  This technique offers the advantage of generating broadband radiation with intensity levels comparable to those of  a standard commercially available zinc telluride source. 

A magnetic heterostructure generating THz radiation, referred to as a spintronic THz emitter (STE), typically consists of a heavy metal (HM) with strong spin-orbit interaction and a ferromagnetic (FM) material. It is well established that a femtosecond laser pulse leads to an ultrafast demagnetization process of the FM layer and the generation of a superdiffusive spin current ($j_s$) \cite{PhysRevLett.105.027203,PhysRevB.108.064427}. The transfer of spin current across the FM/HM interface will via the inverse spin Hall effect (ISHE) \cite{RevModPhys.87.1213} or the inverse Rashba-Edelstein effect (IREE) \cite{Sánchez2013} generate a transient charge current ($j_c$) in the HM layer that emits THz radiation. The critical parameters for the emission of THz radiation are the spin-to-charge conversion efficiency characterized by the spin Hall angle ($\theta_{SHE}$) of the HM, $\theta_{SHE}=j_s/j_c$, and the transparency of the FM/HM interface characterized by the effective spin-mixing conductance ($g_{eff}^{\uparrow\downarrow}$) of the interface \cite{PhysRevB.66.224403,10.1063/5.0022369,PhysRevB.101.024401,Piyush,Rekha}. The effective spin-mixing conductance describes the transfer of spin current across the interface and accounts for spin-backflow and spin-memory loss at the interface, emphasizing the importance of the interface quality for STEs. For example, a high interface roughness or interfacial alloying leads to an enhanced loss of spin angular momentum at the interface \cite{PhysRevLettgupta}. As a result, the interfacial atomic structure and the crystalline quality of the heterostructure play a decisive role in optimizing the THz emission. 
Although interfacial transparency is central, several other factors also affect the emission efficiency, including FM spin polarization, optical absorption in ultrathin layers, film quality , layer thickness and substrates as well as external coupling elements such as silicon lenses and antennas \cite{seifert2016efficient,gupta2021co2feal,PapaioannouBeigang,scheuer2022thz,torosyan2018optimized,nandi2019antenna,Buryakov1}. In ultrathin films, optical absorption is typically limited to ~50\%, motivating structural innovations such as stacked STEs, dielectric cavities, photonic crystal integration, and silicon-based STEs to enhance absorption \cite{Yang2016,feng2018highly,herapath2019impact,APL.120.042404,rouzegar2023broadband}.

Material selection is equally crucial. CoFe alloys are highly attractive FMs due to their larger spin polarization ($\sim$58\%) and higher saturation magnetization compared to elemental Fe or Co \cite{JAP.105.07C916}. Previous THz studies on Co$_x$Fe$_{1-x}$/Pt heterostructures confirmed their usefulness as magnetic layers \cite{APL.120.042404}. However, those investigations relied on polycrystalline films, leaving interfacial and epitaxial effects largely unexplored. In fact, epitaxial CoFe(001) films have been reported to exhibit slightly higher spin polarization than their polycrystalline counterparts \cite{JAP.105.07C916}, suggesting a potential impact on THz emission. In support of this view, epitaxial Pt/Fe and Co/Pt emitters have demonstrated stronger THz signals than their polycrystalline analogs \cite{torosyan2018optimized, PhysRevMaterials.3.084415}, although a somewhat contradictory result was found when comparing epitaxial Fe/Pt and amorphous CoFeB/Pt emitters both showing similar THz signals \cite{seifert2016efficient,kumar2021optical}. This ambiguity highlights the role of the crystalline structure and remains under debate.

Systematic STE studies comparing different choices of HM   \cite{seifert2016efficient} revealed that Pt performs best in terms of  THz yield. Moreover, epitaxial Pt (on Fe) produced a THz amplitude twice as high as compared to partially epitaxial Pt (on Fe), which was attributed to an increase of the electron spin lifetime \cite{nenno2019modification}. To reduce material costs, researchers are actively exploring alternatives to expensive Pt by employing more affordable HM such as W, Ta, and Pt-based alloys \cite{janus2025,shi2026,Yadav1,kumar2021optical}. Among these candidates, W is a particularly intriguing one. It exists in two phases: the stable $\alpha$-W (A2) with low resistivity ($\sim$20 $\mu \Omega$cm) and modest spin Hall angle ($\theta_{SHE} \sim -0.07$), and the metastable $\beta$-W (A15) with high resistivity (150–300 $\mu \Omega$cm) and a giant spin Hall angle ($\theta_{SHE} \sim -0.3$ to $-0.4$) \cite{pai,PhysRevB.96.241105,PhysRevApplied.3.034009}.
Although $\beta$-W exhibits a high spin-Hall angle, its large resistivity weakens the spin-to-charge conversion efficiency and its phase stability remains a concern \cite{Neumann}. The difference arises from the interplay of resistivity and spin Hall conductivity, with theoretical studies suggesting an important role of Dirac fermions and Berry curvature-induced spin Hall conductivity in $\beta$-W \cite{PhysRevB.99.165110,Ye2018}. Interestingly, $\alpha$-W itself is predicted to host topological surface states, particularly in specific crystallographic planes, which may also contribute to spin–charge conversion efficiency \cite{PhysRevB.99.165110,PhysRevB.94.155132,10.1063/5.0009092}.

Despite these advances, systematic studies of epitaxial $\alpha$-W/FM heterostructures with controlled crystallographic orientations are scarce. In particular, how orientation-dependent surface states and interfacial transparency affect spin-current transmission and THz emission remains largely unexplored.

In this work, we investigate the THz emission from epitaxial heterostructures $\alpha$-W/Co$_{60}$Fe$_{40}$ grown on single-crystalline Al$_2$O$_3$($11\bar{2}0$) and MgO($001$) substrates, stabilizing the $\alpha$ -W(110)/CoFe(110) and $\alpha$ -W(001)/CoFe(001) interfaces, respectively. This allows us to directly compare orientation-dependent properties, including the possible role of Dirac-type surface states at $\alpha$-W(110) and $\alpha$-W(001), and to explore how the THz emission efficiency is affected by interfacial strain. Moreover, we show how complementary measurements of reflectance and transmission spectra as well as Fresnel calculations can contribute to a broader understanding of THz generation in spintronic THz emitters, since such analysis helps to distinguish intrinsic properties from extrinsic factors affecting the THz emission.




\section{Results and Discussion}

\begin{figure*}
    \centering
    \includegraphics[width=13cm]{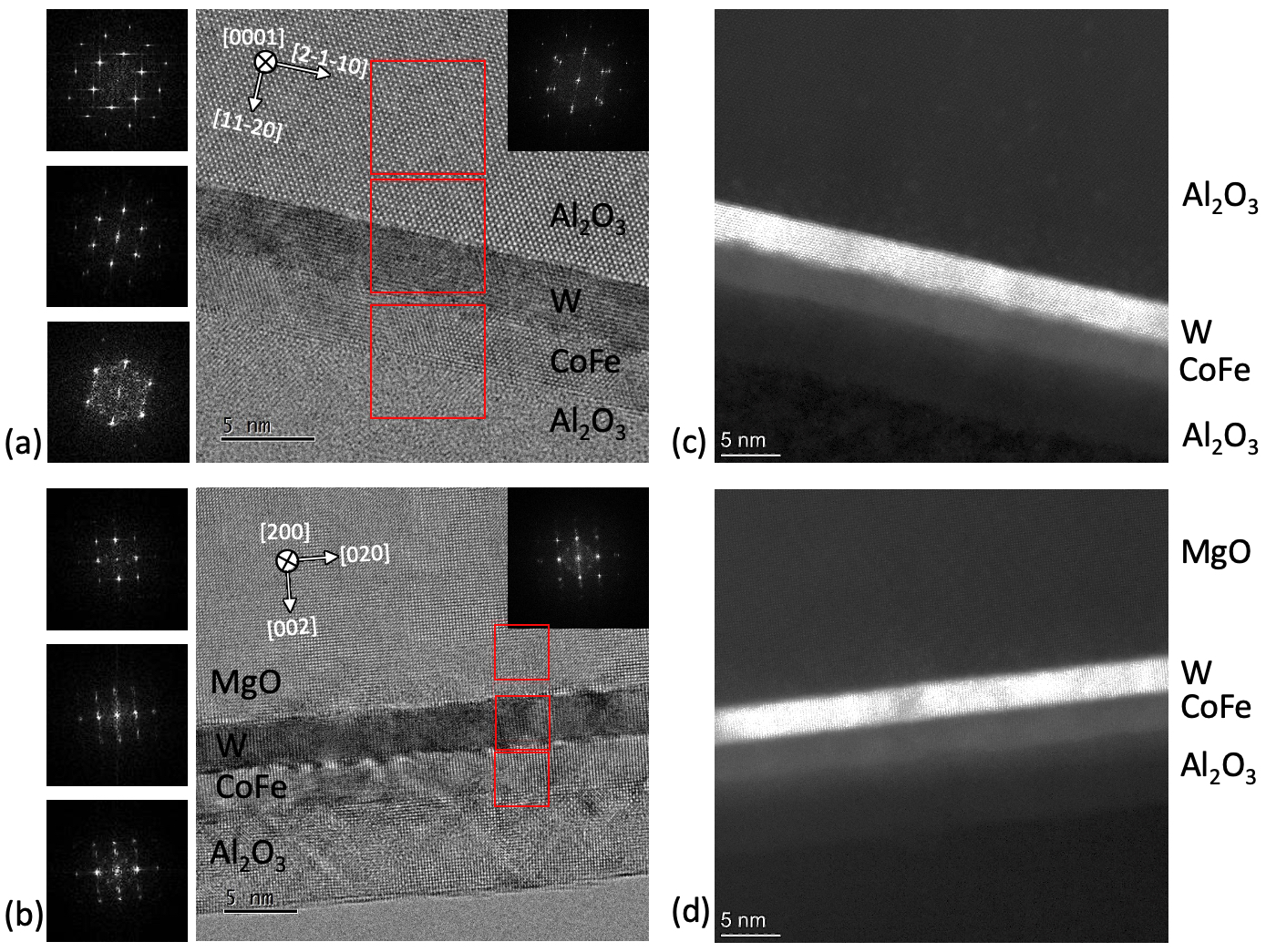}
        \caption{HRTEM and high-resolution HAADF-STEM images of W/CoFe layers on (a,c) Al$_2$O$_3$ and (b,d) MgO substrate, respectively. Inset: overall FFT of the HRTEM image. FFT patterns of selected areas, highlighted by red squares, are also given on the left side of the images.} \label {fig:Figure1}
\end{figure*}

High-Resolution Transmission Electron Microscopy (HRTEM) and high-resolution High-Angle Annular Dark Field Scanning Transmission Electron Microscopy (HAADF-STEM) images of W/CoFe layers are shown in Figure \ref{fig:Figure1}, while HAADF-STEM images and Energy Dispersive X-Ray Spectroscopy (EDS)  maps are presented in Supplementary Material (SM), Fig. S1. Figures \ref{fig:Figure1} (a) and (b) show the HRTEM images of W/CoFe on Al$_2$O$_3$ and MgO substrate, respectively. The asymmetric contrasts of different lattice fringes indicate a slight but detectable crystallographic misorientation from the zone axis. The alignment of the lattice planes in W, CoFe, and the substrates is consistent with the HAADF results. The estimated thicknesses of the W and CoFe layers are $3.2$ nm and $3.4$ nm on the Al$_2$O$_3$ substrate and $3.2$ nm and $3.2$ nm on the MgO substrate, respectively.  The insets in Figure \ref{fig:Figure1} (a) and (b) are the fast Fourier transform (FFT) of the HRTEM images, which are similar to the selected area electron diffraction patterns shown in SM, Figs. S2 and S3. In addition, the FFTs of the selected regions,  highlighted in red squares, show the orientations. All FFTs confirm the epitaxial relationship. 

The contrast of the HAADF-STEM images shown in Figs. \ref{fig:Figure1} (c) and (d), known as Z-contrast, clearly distinguishes the layers, including the Al$_2$O$_3$ capping layer. The EDS maps shown in SM, Fig. S1 confirm that the element distributions are confined to each layer. In the case of the MgO substrate, the interface between the W and CoFe layers is atomically flat, while a few atomic layers intermixing at the interface are observed for the Al$_2$O$_3$ substrate (cf. HAADF-STEM images in Fig. \ref{fig:Figure1} (c) and in SM, Fig. S1 (a)). The difference in W/CoFe interface roughness between the two substrates is also confirmed by the X-ray reflectometry (XRR) results (cf. SM, Tables ST1-ST4). In the case of the Al$_2$O$_3$ substrate, the interface between the W(110) and CoFe(110) layers in the HRTEM image (Fig. \ref{fig:Figure1} (a)) appears smoother than indicated by HAADF-STEM. It is worth noting that the imaging mechanisms of HRTEM are phase and diffraction contrast, which are sensitive to crystallographic orientation and defects but not to the atomic number as in HAADF. 

The interface in the film deposited on MgO exhibits localized bright/dark contrasts (cf. Fig. \ref{fig:Figure1} (b)). These additional contrasts indicate the presence of local strain, where distortion of lattice fringes and edge dislocations is expected. The strain of the film deposited on MgO was estimated using geometric phase analysis (GPA) \cite{HYTCH1998131}, which measures the difference between a reference strain-free region and the region of interest. The former region was selected about 10 nm away from the MgO($001$)/$\alpha$-W($001$) interface. The result of the quantitative strain analysis is presented in SM, where the HRTEM image used for GPA is shown in Fig. S4 (a), strain maps along and perpendicular to the interface are shown in Figs. S4 (b) and (c), respectively, and line profile scans are shown in Figs. S4 (d) and (e). The strain analysis confirms the presence of local strain at the W($001$)/CoFe($001$) interface for the MgO substrate, which can reduce the transparency of the spin current due to spin scattering and spin memory loss \cite{PhysRevLettgupta,PhysRevLett.122.077201}. The corresponding GPA analysis for the film deposited on Al$_{2}$O$_{3}$ is shown in Fig. S5. The variation of strain is about $\pm 1\%$ and is quite different from the film deposited on MgO; no obvious strain gradient is observed in this case.

\begin{figure*}
    \centering
    \includegraphics[width=13 cm]{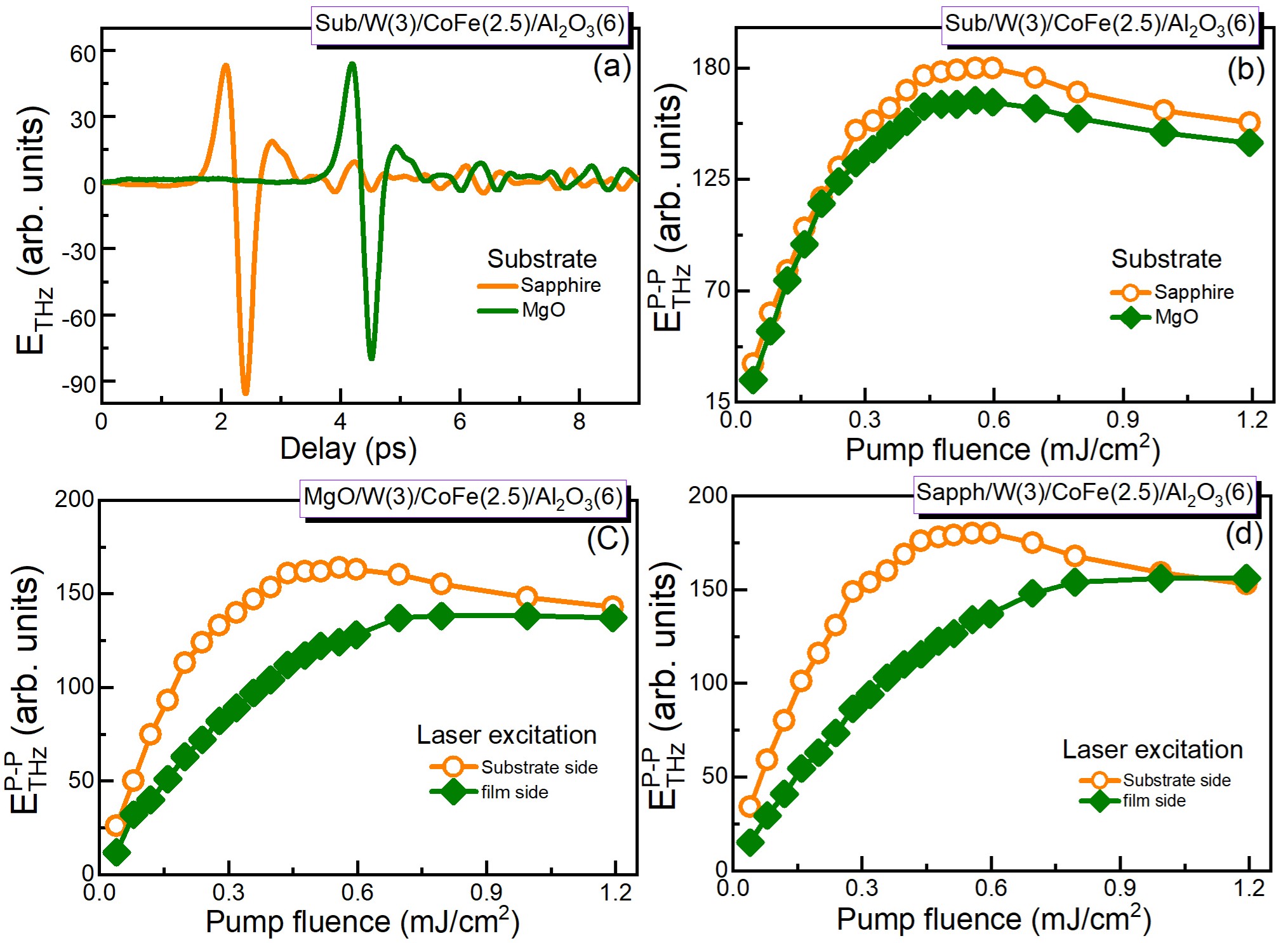}
        \caption{(a) THz electric field waveform in time-domain for substrate/W($3$)/CoFe($2.5$) when pumping from substrate side using $0.3$ mJ$/$cm$^2$ laser fluence (waveforms are horizontally shifted ($\sim$2ps) for clarity). (b) THz electric field peak-to-peak amplitude for substrate/W($3$)/CoFe($2.5$) versus laser fluence. (c) THz electric field peak-to-peak amplitude for MgO/W($3$)/CoFe($2.5$) versus laser fluence when pumping from the substrate and film sides. (d) THz electric field peak-to-peak amplitude for Al$_2$O$_3$/W($3$)/CoFe($2.5$) versus laser fluence when pumping from the substrate and film sides.}\label {fig:Figure2}
\end{figure*}

Figure \ref{fig:Figure2}(a) shows the generated THz electric field in time domain for W/CoFe STEs deposited on Al$_{2}$O$_{3}$~($11\bar{2}0$) and MgO~($001$) substrates when pumping from the substrate side. Note that the polarity of the THz waveform is reversed when reversing the magnetization direction while keeping the pumping side the same (cf. Fig. S6 in SM). Thus, the THz emission is clearly of magnetic origin. Figure \ref{fig:Figure2}(b) shows the THz electric field peak-to-peak amplitude versus laser fluence for the same STEs. As expected, the peak-to-peak amplitude increases linearly at low laser fluence, but the increase slows down above $0.2$ mJ$/$cm$^2$ laser fluence, and the amplitude reaches a maximum at about $0.6$ mJ$/$cm$^2$ laser fluence followed by a weak decrease of the amplitude on further increase of the laser fluence. Similar behavior has previously been observed and can be attributed to spin accumulation and laser-induced heating effects \cite{Kampfrath2013-yt,Huisman2016-ul,Yang2016}. The spin accumulation effect implies that there is an upper limit for the number of spin-polarized electrons that the HM layer can host, while the heating effect induces a sufficiently large increase in spin temperature to weaken the magnetization of the ferromagnetic layer \cite{PhysRevB.85.184301}. This implies that the leveling off of the emitted THz radiation can be explained by both the saturation of the spin accumulation in the W-layer and heating of the spin system, whereas the weak decrease of the emitted THz radiation can only be explained by heating of the spin system and a reduction of the spin polarization in the CoFe-layer. The variation of the peak-to-peak amplitude of the THz electric field with thickness of the W and CoFe layers is reported in SM, Fig. S7. The largest amplitude is obtained for a thickness of W of $2.5$nm and a thickness of CoFe of $3$nm.

Figures \ref{fig:Figure2}(c,d) show, respectively, the THz peak-to-peak electric field amplitude for MgO/W($3$)/CoFe($2.5$) and Al$_2$O$_3$/W($3$)/CoFe($2.5$) versus laser fluence when pumping from the substrate and film sides. For both STEs, the generated THz electric field is significantly larger when pumping from the substrate side. A plausible explanation for this is that more laser fluence is absorbed when pumping from the substrate side and therefore a larger spin current is generated, which will be discussed next. 

\begin{figure*}
    \centering
    \includegraphics[width=13 cm]{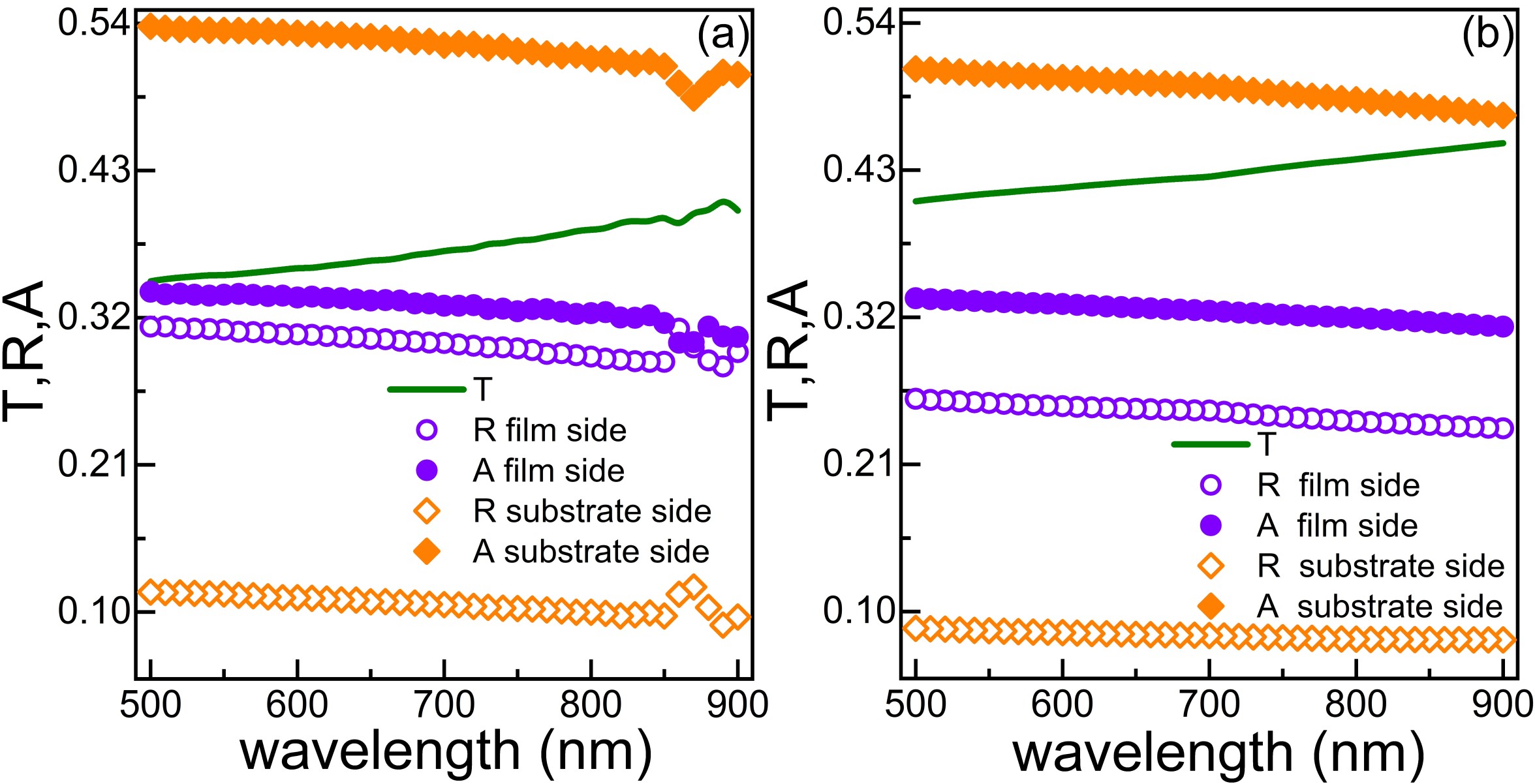}
        \caption{Reflectance, transmittance and absorptance spectra for a smooth, non-scattering sample measured with the film and substrate sides facing the beam. It should be noted that there is no difference in transmittance between the two measurement configurations in this case. (a) Experimental results for Al$_{2}$O$_{3}$/W($3$)/CoFe($3$) and (b) calculated results for Al$_{2}$O$_{3}$/W($3$)/CoFe($3$).}
        \label {fig:Figure3}
\end{figure*}

There is a considerable difference in optical properties between the two measurement orientations, as light beams reflected from the two surfaces of the substrate are non-coherent and are not subject to interference effects. The aluminum-oxide cap on the film side, on the other hand, is very thin and thus interference occurs between multiply reflected beams. The surface texture of the back surface of the substrate also causes light to be scattered in different ways in the front and back configurations. These two effects result in the fact that the transmittance and reflectance spectra are different for the two measurement orientations. The difference in absorptance caused by interference is illustrated in Fig. \ref{fig:Figure3}(a), where the reflectance, transmittance, and absorptance spectra of the W(3)/CoFe(3) emitter are shown. This sample was prepared on a double-sided polished Al$_{2}$O$_{3}$~($11\bar{2}0$) substrate, implying that there is no surface texture on the back surface of the substrate. It can be seen that for this sample, the reflectance is quite different between the light incident from the substrate and film sides, whereas the transmittance is the same. In both orientations, the diffuse signals are close to zero. This means that the absorptance is higher for light incident from the substrate side, which in turn implies that one can expect a stronger spin current $j_s$ to be generated for this orientation.

For the other STEs the surface texture on the back surface of the substrate causes diffuse (scattered) signals, which complicates the situation. In these cases, the transmittance is not the same from the two sides, and both reflectance and transmittance differ. This can be seen in SM, Fig. S9 showing the reflectance, transmittance and absorptance spectra for W($3$)/CoFe($2.5$) and W($2.5$)/CoFe($3$) grown on single-side polished substrates of Al$_{2}$O$_{3}$ and MgO, respectively. However, it can be seen that the absorptance for all STEs is higher for light incident from the substrate side.

A problem with the results from the calculations is that these do not take light scattering into account, but the results indicate that the trends with varied input parameters are consistent with our experimental results. The sample with a smooth back surface can be used to compare the experimental and calculated results with no scattering involved. This is shown in Fig. \ref{fig:Figure3}(b). It can be seen that the trends with the difference in reflectance and absorptance when light is incident on the film and substrate sides of the sample are consistent. As explained above, the contribution to the absorptance from the substrate is marginal. Considering the extremely thin layers used in the experiments, the agreement between the experimental and calculated spectra in Figs. \ref{fig:Figure3} (a) and (b) is surprisingly good and very encouraging.

The property of interest in this investigation is the absorptance within the film stack. Neither the experimental results nor the calculated results provide information about exactly where in the film stack absorption occurs. However, it is obvious that the absorption mechanism is within the two metallic layers. Because of the similarity of the optical constants of W, Co, and Fe, the absorption is likely to be similar in the two metallic layers (W and CoFe). Both the calculated results and the experimental results indicate that the absorptance is higher when light is incident from the substrate side. The absorptance also increases with the film thickness of the CoFe layer, which is also as expected. These results are shown in SM, Fig. S10 for the measured and calculated absorptance. The trend is the same for all results, but there is a difference as the experimental results take light scattering into account, whereas the calculated spectra assume ideally flat interfaces.

Figure \ref{fig:Figure2} also indicates that the emitted THz electric field is about $10\%$ larger for the STE grown on the Al$_{2}$O$_{3}$~($11\bar{2}0$) substrate. The absorptance spectrum for the STE grown on a single-side polished MgO (cf. Fig. S9(b)) indicates a slightly larger absorptance value compared to an STE grown on a single-side polished Al$_2$O$_3$ substrate. This indicates that the $10\%$ larger THz emission for the emitter deposited on Al$_2$O$_3$ can best be explained by the higher spin-to-charge conversion efficiency of the W(110) layer and/or the higher spin transparency of the W/CoFe interface.


It is tempting to attribute this difference to the Dirac-type surface states shown to exist in $\alpha$-W(110) \cite{PhysRevB.94.155132}. To gain a deeper understanding, we performed first-principles calculations for $\alpha$-W(001), investigating both the bulk electronic band structure and the W(001) surface. The electronic structure of the surface of W(110) has been previously reported by the authors in Refs. \cite{PhysRevB.94.155132,Mirhosseini2013}. In analogy to the Dirac-type surface states (DSS) observed in that work, we identified two DSS also in the (001) crystallographic plane (see Fig. \ref{fig:Figure4}), located at the $\Gamma$ point at $E_F - 0.58$ eV and the X point at $E_F - 0.66$ eV. Unlike the W(110) surface, the DSS on W(001) exhibits linear dispersion only in the immediate vicinity of the $\Gamma$ point. Orbital decomposition of the spectral function of the top-layer reveals a dominant $d_{z^2}$ character, consistent with the findings of W(110).

The spin texture exhibits Rashba-type characteristics governed by point group symmetry and time-reversal symmetry. Specifically, the spin lies in-plane and is oriented perpendicular to the wave vector $\vec{k}$, the so-called Rashba component. The time-reversal symmetry requires that the reversal of $\vec{k}$ also reverses the spin direction. Near the Dirac point, the spin polarization reaches approximately 30\% in absolute terms, lower than that observed in topological insulators such as Bi$_2$Te$_3$ (about 60\% \cite{henk2012complex}). The Chern number, which reflects the number of topological surface states, is identical for the W(001) and W(110) surfaces. Consequently, the Berry curvature extracted from the spectral function~\cite{PhysRevB.94.155132} is expected to be similar for both orientations. However, because the spin Hall effect (SHE) is proportional to the integrated Berry curvature, drawing firm conclusions about which surface exhibits a stronger SHE requires more comprehensive simulations. In contrast, the in-plane conductivity of these surfaces is primarily governed by the density of states at the Fermi level. Due to the presence of band touching between $\overline{\Gamma}$ and $\overline{X}$ in W(001), this surface is expected to exhibit higher conductivity than W(110) (see Fig.~ 1 of Ref. \cite{Mirhosseini2013}).

\begin{figure*}
    \includegraphics[width=12cm]{Figure4.jpg}
    \caption{Electronic band structure with marked Dirac-type surface states (DSS) of $\alpha$-W (001) crystallographic planes. a) Left and right panel shows the bulk and surface band structure, respectively. b) Spin polarisation (red - spin-up; blue - spin-down) of the surface states. The Brillouin zone of a bcc lattice and its projection onto the $(001)$ surface is shown as inset.}\label {fig:Figure4}
\end{figure*}

The contribution of Dirac states - and more broadly, of the Berry curvature - to the off-diagonal conductivity component $\sigma_{xy}$ is well established~\cite{PhysRevB.99.235113}. The Hall angle is defined as the ratio $\sigma_{xy}/\sigma_{xx}$. These transport coefficients are often calculated from bulk electronic structures~\cite{science.1058161,PhysRevB.95.075128}, but similar methods can be applied in slab geometries, as demonstrated for the anomalous Hall effect in Ref. \cite{PhysRevB.98.115108}. While the diagonal components of the conductivity tensor arise from Fermi-surface states, the off-diagonal components involve Fermi-sea contributions, implying that lower-lying Dirac states also play a role. Since the two surfaces of a slab may differ significantly in their band structure - as shown in this work - surface-resolved differences in $\sigma_{xy}$, $\sigma_{xx}$, and therefore the spin Hall angle is expected.

Regarding spin Hall conductivity, the spin-current operator is a central quantity \cite{PhysRevB.99.235113}, especially in systems featuring strongly spin-polarized states such as Dirac bands. For surfaces lacking Dirac-type states, we do not anticipate large variations in the spin Hall angle between different surface orientations. Additionally, finite conductivity requires metallic behavior. Therefore, it is reasonable to suggest that Dirac surface states in elemental transition metals such as Ta, Mo, and Nb~\cite{PhysRevB.94.155132} could also contribute to the spin Hall angle. However, this hypothesis requires further investigation through detailed numerical simulations.

Although an explanation based on a difference in the electronic structure of the surface cannot be ruled out, the results of the present calculations are not conclusive in this respect. Another possible explanation for the difference in the amplitude of the THz electric field is connected to the transparency of the $\alpha$ -W / CoFe interface characterized by the effective spin-mixing conductance. The influence of interface intermixing on spin-mixing conductance has been quantitatively studied for Pt/Py (Py = Permalloy) bilayers using first-principles calculations \cite{Zhang2011}. The interface intermixing was modeled as a scattering region consisting of composite Pt$_{1-x}$Py$_x$ and Pt$_x$Py$_{1-x}$ layers for the ﬁrst Pt and Py atomic layers at the interface. Compared with the spin-mixing conductance for an ideally clean and flat atomic interface, corresponding to $x=0$, it was found that the spin-mixing conductance is enhanced in the presence of intermixing of the interface. Moreover, experimental support for this has been obtained for Co/Pt bilayers \cite{PhysRevMaterials.3.084415}. In this work, an ultrathin layer of the Co$_x$Pt$_{1-x}$ compound was introduced between the Co and Pt layers to study the effect of interface intermixing, clearly revealing that interface intermixing improves THz emission. Based on the information obtained from the high-resolution HAADF-STEM images, we argue that an effect contributing to the enhanced THz electric field amplitude for the STE grown on the Al$_{2}$O$_{3}$~($11\bar{2}0$) substrate could be the intermixing at the interface consisting of a few atomic composite layers as described in the first-principles calculations \cite{Zhang2011}. Another contributing effect to the difference in THz emission is the presence of local strain in the $\alpha$-W$(001)$ layer grown on single crystalline MgO($001$), which due to spin memory loss will reduce the spin current. Considering the high crystalline quality of the bilayers studied here, we are not considering the effect of strongly disordered interfaces on the spin-mixing conductance and THz emission. 

\begin{figure*}
    \centering
    \includegraphics[width=13 cm]{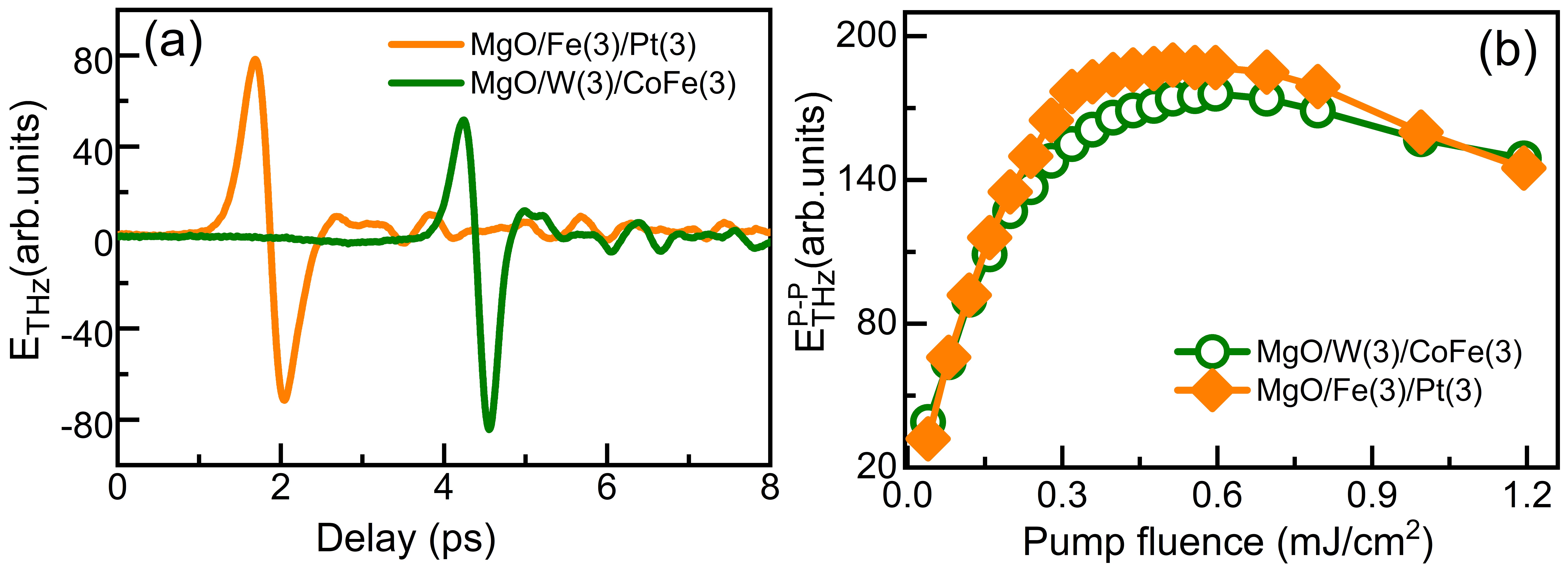}
        \caption{(a) THz electric field waveform in time-domain for MgO/W($3$)/CoFe($3$) and MgO/Fe($3$)/Pt($3$) when pumping from the substrate side using $0.24$ mJ$/$cm$^2$ laser fluence (waveforms are horizontally shifted by $2.5$ ps for clarity). (b) THz electric field peak-to-peak amplitude for MgO/W($3$)/CoFe($3$) and MgO/Fe($3$)/Pt($3$) when pumping from the substrate side versus laser fluence.} \label {fig:Figure5}
\end{figure*}

It is interesting to compare the $\alpha$-W / Co$_{60}$Fe$_{40}$ STE investigated here with a high-performance STE such as Fe/Pt \cite{Yang2016}. We therefore deposited a MgO/Fe(3)/Pt(3) STE for comparison with $\alpha$-W(001) based STEs. Figures \ref{fig:Figure5} (a) and (b) show the THz electric field waveform in the time domain and the peak-to-peak amplitude versus laser power for MgO/W($3$)/CoFe($3$) and MgO/Fe($3$)/Pt($3$), respectively. The variation of the THz emission amplitude of Fe/Pt for different thicknesses of the Pt layer is presented  in Fig. S8 in SM. Earlier reports have shown approximately twofold higher THz emission amplitude for Pt/Co compared with W/Co  and a threefold higher THz emission amplitude comparing Pt/NiO with W/NiO \cite{Wu2017,qiu2021ultrafast}. 
The peak-to-peak THz amplitude obtained from epitaxial $\alpha$-W /Co$_{60}$Fe$_{40}$ grown on  MgO is comparable to the amplitude measured on epitaxial Fe/Pt films of similar thickness grown on the same substrate under identical excitation conditions. A comparison of the effect of structural order on the THz emission amplitude across various material systems, including our results, is presented in SM, Table  ST5. 

\section{Methods}

\subsection{Growth}

Magnetic heterostructures consisting of  $\alpha$-W/Co$_{60}$Fe$_{40}$ (henceforth referred to as W/CoFe) bilayers with different layer thicknesses ranging from $2$ to $4$ nm were deposited on single crystalline Al$_{2}$O$_{3}$~($11\bar{2}0$) and MgO~($001$) substrates (both 10$\times$10~mm$^2$ with a thickness of $0.5$ mm) at floating potential, using direct current (dc) magnetron sputtering. Except for the W($t_{W}$)/CoFe(3) series of samples ($t_{W}$ is the thickness of the W-layer; numbers in parentheses refer to layer thickness in nanometers), which were deposited on double-side polished Al$_{2}$O$_{3}$~($11\bar{2}0$), the films were deposited on single-side polished substrates. Before deposition, the substrates were cleaned in acetone and 2-propanol using ultrasonic agitation for 120~s each. This was followed by vacuum annealing at 873(2)~K for 1 hour. The base pressure of the growth chamber was below 5$\times$10$^{-7}$~Pa. To prevent surface oxidation of the films, the samples were capped at ambient temperature ($<$~313(2)~K) with Al$_2$O$_3$ (nominal thickness 6 nm) using radio frequency (rf) magnetron sputtering. The depositions were carried out in Ar atmosphere at a pressure of 3$\times$10$^{-2}$~Pa (gas purity $\geq$~99.999~\%) from an elemental W (25~W, dc) target, and CoFe (13~W, dc) and Al$_{2}$O$_{3}$ (90~W, rf) compound targets. The targets were cleaned by sputtering against closed shutters for at least 60~s prior to each deposition. The target-to-substrate distance in the deposition chamber was around 0.2~m. The deposition rates (W: 0.253~{\AA}/s, Al: 0.30~{\AA}/s, CoFe: 0.10~{\AA}/s, Al$_{2}$O$_{3}$: 0.03~{\AA}/s) were calibrated prior to growth using X-ray reflectivity. The W growth temperature was optimized with respect to the W layering and crystal quality, yielding 843($\pm 2$)~K for single W layers. For the W/CoFe bilayers, W and CoFe were grown at 843($\pm 2$)~K and 573($\pm 2$)~K, respectively. Finally, in order to ensure uniform thickness, the substrate holder was rotated during deposition.

\subsection{Characterization}

A THz time-domain spectrometer was employed to measure the THz emission from the W/CoFe heterostructures  \cite{gupta2021co2feal,gupta2021strain}. This spectrometer utilized a Spectra-Physics Tsunami laser source, which generated pulses with a duration of $\sim 55$ fs (bandwidth $\sim 12$ nm, central wavelength $\sim 800$ nm, and maximum output energy $\sim 10$ nJ) at a repetition rate of 80 MHz.  A low-temperature gallium arsenide photoconductive dipole antenna with a $\sim 4$ $\mu$m gap was used as a detector for the THz pulses. A probe beam with an average laser power of 10 mW was used for the detection and a static in-plane magnetic field of $\sim 85$ mT was used to saturate the magnetization of the W/CoFe films. The recorded THz signals correspond to averages of 500 detected THz spectra obtained within one minute of measurement time. 

Transmission electron microscopy (TEM) and scanning TEM (STEM) were performed in a double aberration-corrected Themis Z (Thermo Fisher), which was operated at 300 kV and equipped with a low-background double-tilt holder and a Super-X EDS detector. The aberrations were corrected up to 5th order. STEM imaging and EDS acquisition and analyses were acquired using a Thermo Fisher Velox. The convergent angle of the probe and the collection angle for high-angle annular dark-field (HAADF) imaging were 16 and 63-200 mrad, respectively. TEM and selected-area electron diffraction were recorded using a Gatan OneView camera. 

X-ray reflectometry (XRR) and diffraction (XRD) were carried out on a Bede D1 diffractometer equipped with a Cu $K_{\alpha_1}$ x-ray source operated at 35~mA and 50~kV. A circular mask (diameter: 0.005~m) and an incidence and a detector slit (both 0.0005~m) were used. For monochromatizing the beam by reducing the CuK$\beta$ and CuK$\alpha_2$ radiation, the setup included a G\"obel mirror and a 2-bounce-crystal on the incidence side. The x-rays were detected with a Bede EDRc x-ray detector. The XRR results for the STEs studied here are presented in SM, Tables ST1-ST4. The XRD results, described elsewhere \cite{ravensburg2023epitaxy}, show that $\alpha$-W grows epitaxially in the [$110$] growth direction on Al$_{2}$O$_{3}$~($11\bar{2}0$) and in the [$001$] growth direction on MgO ($001$). The lattice mismatch using different substrates may cause strain and misfit dislocations in the layers. The epitaxial growth of $\alpha$-W on both substrates has been described in detail in Ref. \cite{ravensburg2023epitaxy}. Due to a few percent of the lattice mismatch between MgO and $\alpha$-W(001), there is approximately 2\% of out-of-plane strain in $\alpha$-W(001) \cite{ravensburg2023epitaxy}. The lattice mismatch is greater in the case of Al$_{2}$O$_{3}$($11\bar{2}0$)/$\alpha$-W(110). However, due to the semi-coherent growth mode that includes the introduction of misfit dislocations in the immediate vicinity of the substrate/film interface, the result is a nearly strain-free layer $\alpha$-W \cite{ravensburg2023epitaxy}. Comparing the two substrates, it can be concluded that $\alpha$-W grows epitaxially on both substrates with high crystal quality, but that the crystal quality is better for $\alpha$-W(110) grown on Al$_{2}$O$_{3}$.

A Quantum Design magnetic property measurement system was used to assess the magnetic properties of the samples; results from room temperature magnetization versus in-plane magnetic field measurements are presented in SM, Fig. S11. An AIT CMT-SR2000N 4-point probe measurement system was used to measure the sheet resistance of $\alpha$-W films with thicknesses in the range $6-100$ nm. The low resistivity of the measured samples (cf. Fig. S12 in SM) confirms the $\alpha$-phase for the epitaxial W layers. 

In an attempt to determine the optical absorptance of the investigated film stacks, the reflectance spectra ($R$) and transmittance spectra ($T$) were measured using an integrated sphere, a Perkin Elmer Lambda 900 double beam spectrophotometer equipped with a $15$ cm spectralon-coated sphere. The scattered and regular/specular light signals entering the sphere have to be corrected in different ways for the sphere wall reflectance, and therefore both total and scattered spectra were recorded in both reflectance and transmittance modes \cite{FENDLEY1985281,ClarkeCompton,Roos:88}. The absorptance was then calculated as $A=1-R-T$. The calculated absorptance is the total absorptance of the film stack. Schematic illustrations of the sphere geometry in reflectance and transmittance modes are shown in SM, Fig. S13. 

Standard Fresnel calculations were performed for film stacks with different compositions and thickness \cite{BornMax1975Poo,Macleod1969}. The film stack configuration is illustrated in SM, Fig. S14. Such calculations are to some extent uncertain because the included films  are very thin and the optical constants are slightly uncertain. It is well known that the optical constants of very thin films may differ from the optical constants of thick films or bulk materials. Calculations are useful, however, as it is simple and straightforward to vary the composition of the film stack and to get an indication of how the film thickness influences the optical properties of the samples. The optical constants for the different layers and substrates used were taken from the literature \cite{Palik:1991,Weaver}. The optical constant for the CoFe film was taken as the average value of the constants for Co and Fe. 

The optical properties of the bare substrates of thickness $0.5$ mm were also measured. Both substrates were single-side polished, i.e. the front surface was polished and smooth, while the back surface had some surface roughness causing light scattering. It is a well-known fact that when one surface is causing light scattering, some of the scattered light is coupled into the substrate through total internal reflection and escapes out of the substrate through the edges. This means that it is not possible to separate the absorbed light from the lost scattered light. However, when taking the absorptance as $1-R-T$, which is a standard procedure, an upper limit of the absorptance is obtained. The extinction coefficients for the Al$_2$O$_3$ and MgO substrates are very low and the internal absorptance is expected to be low. The measured values of $1-R-T$ for both sides of the two substrates were between 2 and 7 \%, which means that the actual absorptance from the substrates is  only a few percent at most. Because the absorption in the metallic films used for the THz emission is much higher than this, it was concluded that the possible absorption in the substrates will not significantly affect the results and could be ignored.

The spectral function of the surface band structure was simulated by the density functional relativistic multiple-scattering theory \cite{weinberger1990electron} as formulated in the Korringa-Kohn-Rostoker (KKR) approach \cite{zabloudil2005electron,henk2002thinfilm} and implemented in the CAHMD code \cite{cahmd}. Relativistic effects are fully taken into account by solving the Dirac equation. The surface was modeled by a semi-infinite geometry within the layer-KKR scheme \cite{henk2002thinfilm} which excludes finite size effects and is well suited for semi-infinite systems (e.g., surfaces and interfaces). The spectral density $n_{i\alpha}(E,\boldsymbol{k}_\parallel)$, i.e., the energy- and wavevector-resolved local density of states for a site $\alpha$ in layer $i$, is computed from the site-resolved Green function $G_{i\alpha,i\alpha}(E + i\eta,\boldsymbol{k}_\parallel)$,

\begin{align}
    n_{i\alpha}(E,\boldsymbol{k}_\parallel)=-\frac{1}{\pi}\text{Im} \text{Tr} G_{i\alpha,i\alpha}(E + i\eta,\boldsymbol{k}_\parallel) \:\:\: .
\end{align}

$\eta$ is a small offset from the energy axis leading to a broadening of the spectral density; typically $\eta= 0.01 $eV. The spectral density can be further decomposed with respect to spin polarization, thus allowing for a detailed characterization of the electronic states. In order to account for surface relaxations, we relaxed the surface layer by $-3\%$, similar to the proposed surface layer relaxation in W(110) discussed in Refs. \cite{PhysRevB.94.155132,Mirhosseini2013}.

\section{Conclusions}\label{sec4}

In summary, we provide results for THz emission in epitaxial $\alpha$-W/Co$_{60}$Fe$_{40}$ spintronic emitters. Two types of epitaxial emitters have been investigated; $\alpha$-W$(110)$/Co$_{60}$Fe$_{40}(110)$ and $\alpha$-W$(001)$/Co$_{60}$Fe$_{40}(001)$ deposited on single crystalline Al$_{2}$O$_{3}$($11\bar{2}0$) and MgO($001$) substrates, respectively. The generated THz radiation is about $10\%$ larger for the $\alpha$-W$(110)$/Co$_{60}$Fe$_{40}(110)$ emitter. Even if an explanation based on a difference in surface electronic structure for the $\alpha$-W$(110)$ and $\alpha$-W$(001)$ surfaces can not be ruled out, the results of the calculations presented here are not conclusive in this respect. A more likely explanation is that the $\alpha$-W$(110)$/Co$_{60}$Fe$_{40}(110)$ interface is more transparent to the spin current due to the presence of \AA ngstr\" om-scale interface intermixing at the W/CoFe interface. First-principles calculations show that the spin-mixing conductance is enhanced in the presence of interface intermixing  \cite{Zhang2011}. Another contributing effect to the difference in THz emission is the presence of local strain in the $\alpha$-W$(001)$ layer grown on single crystalline MgO($001$), which due to spin scattering and spin memory loss will reduce the spin current. Our results also reveal that the generation of THz radiation is larger when pumping with the laser light from the substrate side. Measurements of reflectance and transmittance spectra as well as Fresnel calculations for the studied film stacks show that this can be explained by a larger part of the laser light being absorbed in the ferromagnetic Co$_{60}$Fe$_{40}$ layer in this measurement configuration due to interference effects. A comparison with the reference Fe/Pt STE, which is known to provide one of the highest THz signals \cite{PapaioannouBeigang}, shows that the emitted W/CoFe THz electric field amplitude is of similar magnitude. \\

\noindent \textbf{Data availability} \par
\noindent The data presented in the current study are available from the corresponding authors on reasonable request.

\bibliographystyle{unsrt}
\bibliography{sn-bibliography}

\noindent \textbf{Funding} \par 
\noindent This work is supported by the Swedish Research Council (grant numbers 2023-06605, 2023-04239, 2021-04658, 2019-03666, and 2019-03581), Olle Engkvists Stiftelse (grant number 182–0365), the Deutsche Forschungsgemeinschaft (DFG, German Research Foundation) -- Project-ID 328545488 -- TRR~227, project~B04, the Swedish Foundation for Strategic Research (grant number ITM17-0301) and the Royal Swedish Academy of Sciences (project no. PH2025-0043). \\

\noindent \textbf{Author contributions} \par
\noindent VM, RB, RG, VK, and PS contributed to the conception and design of the study. Material preparation, data collection and analysis were performed by VM, RB, ALR, AR, CWT and PS. Theoretical calculations were performed by RM, DT and JH. PS supervised the project. The first draft of the manuscript was written by PS with contributions from VM, DT, AR and CWT and all authors commented on previous versions of the manuscript. All authors read and approved the final manuscript. \\

\noindent \textbf{Declaration of conflicting interest} \par
\noindent The author(s) declared no potential conflicts of interest with respect to the research, authorship, and/or publication of this article. \\

\noindent \textbf{Supporting Information} \par
\noindent Supporting Information is available from the Nature Scientific Reports or from the author.

\end{document}


\title[Article Title]{Terahertz emission from \texorpdfstring{$\alpha$}{}-W/CoFe epitaxial spintronic emitters}

\author*[1,2]{\fnm{Venkatesh} \sur{Mottamchetty}}\email{venkatesh769@gmail.com}
\author[1]{\fnm{Rimantas} \sur{Brucas}}\email{rimantas.brucas@angstrom.uu.se}
\author[2]{\fnm{Anna L.} \sur{Ravensburg}}\email{anna.ravensburg@gmail.com}
\author[3]{\fnm{Renan} \sur{Maciel}}\email{renanm@kth.se}
\author[4]{\fnm{Danny} \sur{Thonig}}\email{Danny.Thonig@oru.se}
\author[5]{\fnm{J\"urgen} \sur{Henk}}\email{juergen.henk@physik.uni-halle.de}
\author[6,7]{\fnm{Rahul} \sur{Gupta}}\email{rahulguptaphy@gmail.com}
\author[1]{\fnm{Arne} \sur{Roos}}\email{Arne.Roos@angstrom.uu.se}
\author[8]{\fnm{Cheuk} \sur{Wai Tai}}\email{cheuk-wai.tai@mmk.su.se}
\author[2]{\fnm{Vassilios} \sur{Kapaklis}}\email{vassilios.kapaklis@physics.uu.se}
\author*[1]{\fnm{Peter} \sur{Svedlindh}}\email{peter.svedlindh@angstrom.uu.se}

\affil[1]{\orgdiv{Department of Materials Science and Engineering}, \orgname{Uppsala University}, \orgaddress{\street{Box 35}, \city{Uppsala}, \postcode{75103}, \country{Sweden}}}

\affil[2]{\orgdiv{Department of Physics and Astronomy}, \orgname{Uppsala University}, \orgaddress{\street{Box 516}, \city{Uppsala}, \postcode{75120}, \country{Sweden}}}

\affil[3]{\orgdiv{Department of Applied Physics}, \orgname{Royal Institute of Technology (KTH)}, \city{Stockholm}, \country{Sweden}}

\affil[4]{\orgdiv{School of Science and Technology}, \orgname{\"Orebro University}, \city{\"Orebro}, \postcode{70182}, \country{Sweden}}

\affil[5]{\orgdiv{Institut f\"ur Physik}, \orgname{Martin Luther University Halle-Wittenberg}, \city{Halle}, \postcode{06099}, \country{Germany}}

\affil[6]{\orgdiv{Department of Physics}, \orgname{University of Gothenburg}, \city{Gothenburg}, \postcode{33620}, \country{Sweden}}

\affil[7]{\orgdiv{Department of Physics}, \orgname{Indian Institute of Technology (Indian School of Mines) Dhanbad}, \city{Dhanbad}, \postcode{826004}, \country{India}}

\affil[8]{\orgdiv{Department of Materials and Environmental Chemistry}, \orgname{Stockholm University}, \city{Stockholm}, \postcode{10691}, \country{Sweden}}

\maketitle
\clearpage
\vspace{\baselineskip}
The Supplementary Material provides HAADF-STEM, EDS maps and electron diffraction patterns (Figs. S1, S2 and S3); strain analysis using geometric phase analysis (Fig. S4 and S5); results from THz measurements (Figs. S6, S7 and S8); experimental and calculated reflectance, transmittance, absorption results (Figs. S9 and S10; magnetization and resistivity results (Figs. S11 and S12); schematic illustration of the sphere geometry in reflectance and transmittance modes and film stack configuration (Figs. S13 and S14); X-Ray reflectivity results (Tables S1-S4); comparison of structural order effects on THz emission in various STE material systems (Table S5). 

\section{HAADF-STEM, EDS maps and electron diffraction patterns} 

\begin{figure}[h!]
   \centering
   \includegraphics[width=10.1cm]{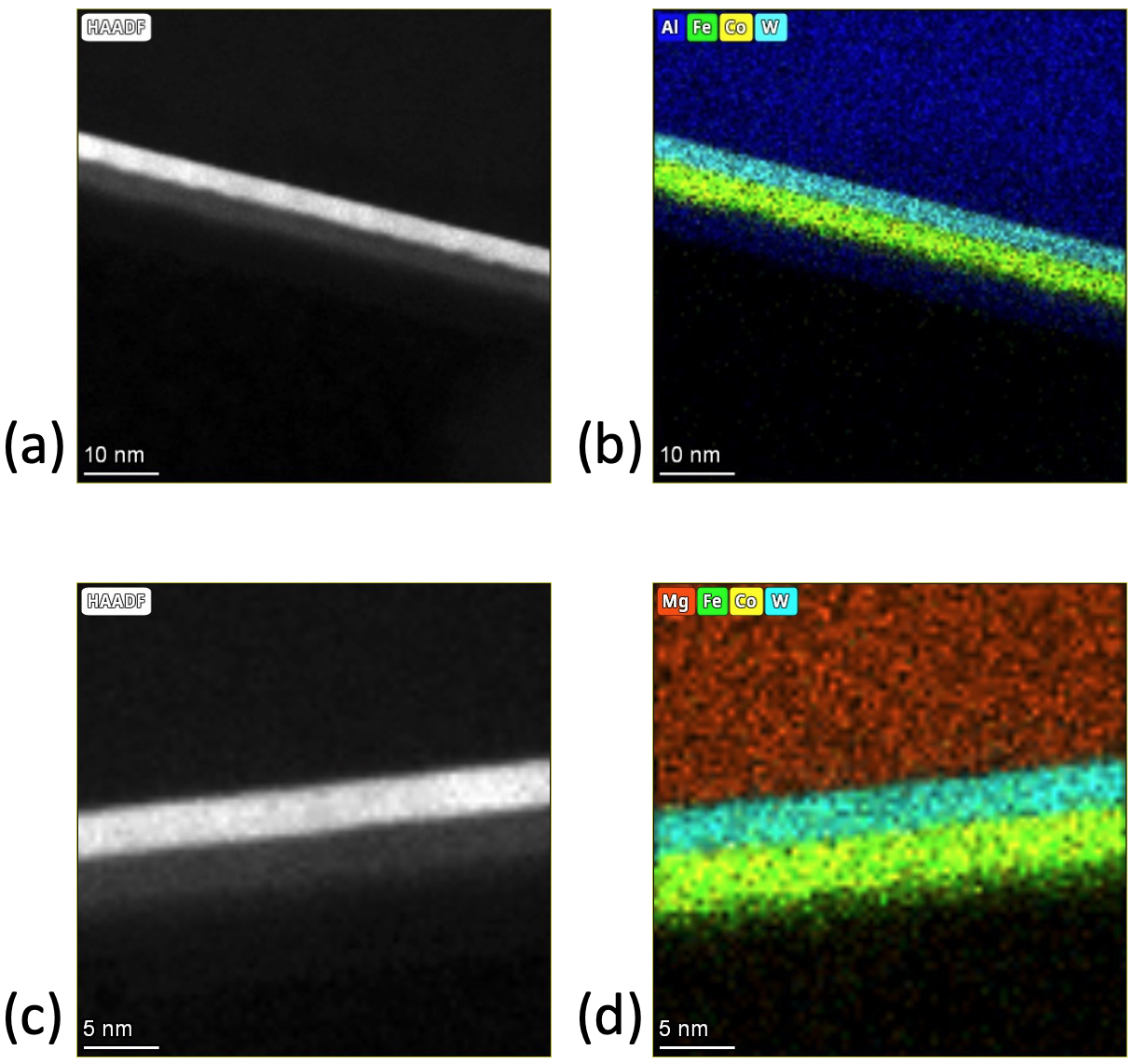}
       \caption{(a) and (c) HAADF-STEM images of W/CoFe 
layers on Al$_2$O$_3$ and MgO substrate, respectively. (b) and (d) Overlay of the elements for each sample. Oxygen was omitted in order to present the elements clearly.} \label {fig:FigureS1}
\end{figure}

\begin{figure}[!ht]
        \centering
 \includegraphics[width=10.5cm]{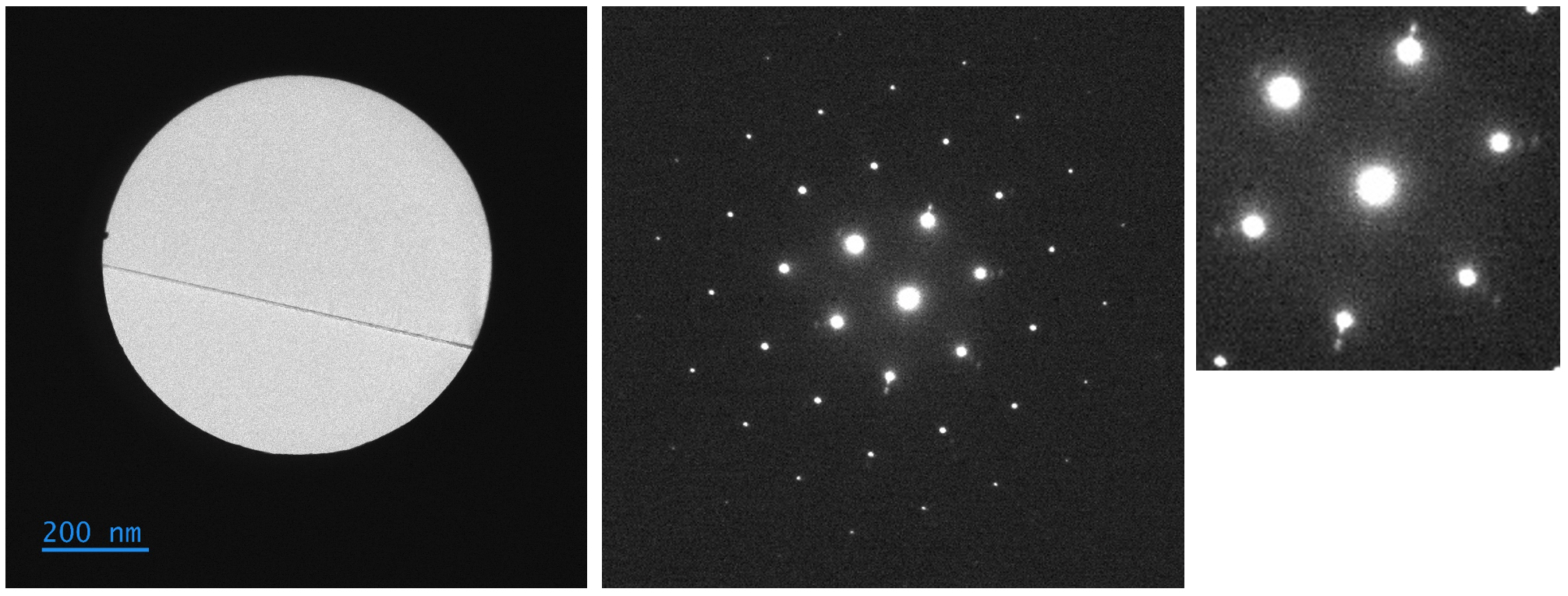}
    \caption{Left: TEM image of W/CoFe layers on Al$_2$O$_3$ taken at lower magnification. Middle: Electron diffraction  pattern of the selected area. Right: Enlarged center region of of the electron diffraction pattern.}
    \label{fig:FigureS2}
\end{figure}

\begin{figure}[!ht]
        \centering
\includegraphics[width=12cm]{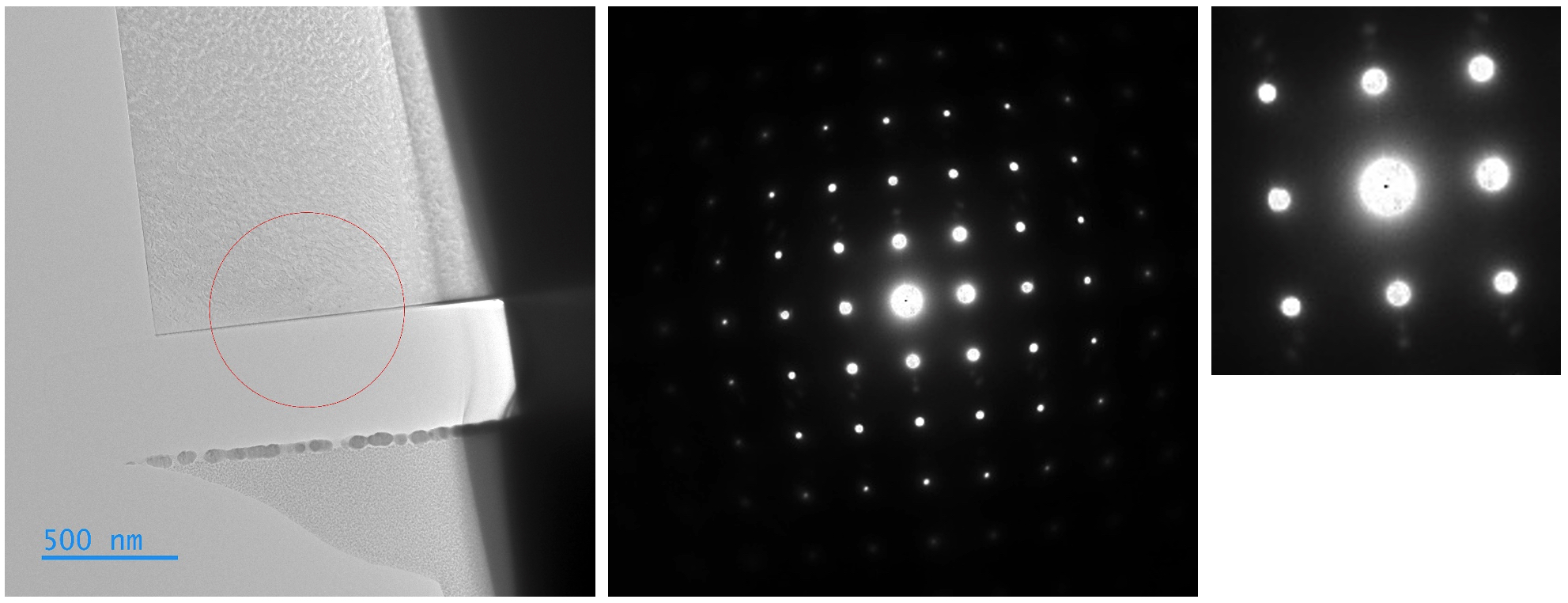}
    \caption{Left: TEM image of W/CoFe layers on MgO taken at lower magnification. The selected area is indicated by a circle. Middle: Electron diffraction  pattern of the selected area. Right: Enlarged center region of of the electron diffraction pattern.}
    \label{fig:FigureS3}
\end{figure}

\begin{figure}[!ht]
        \centering
\includegraphics[width=12cm]{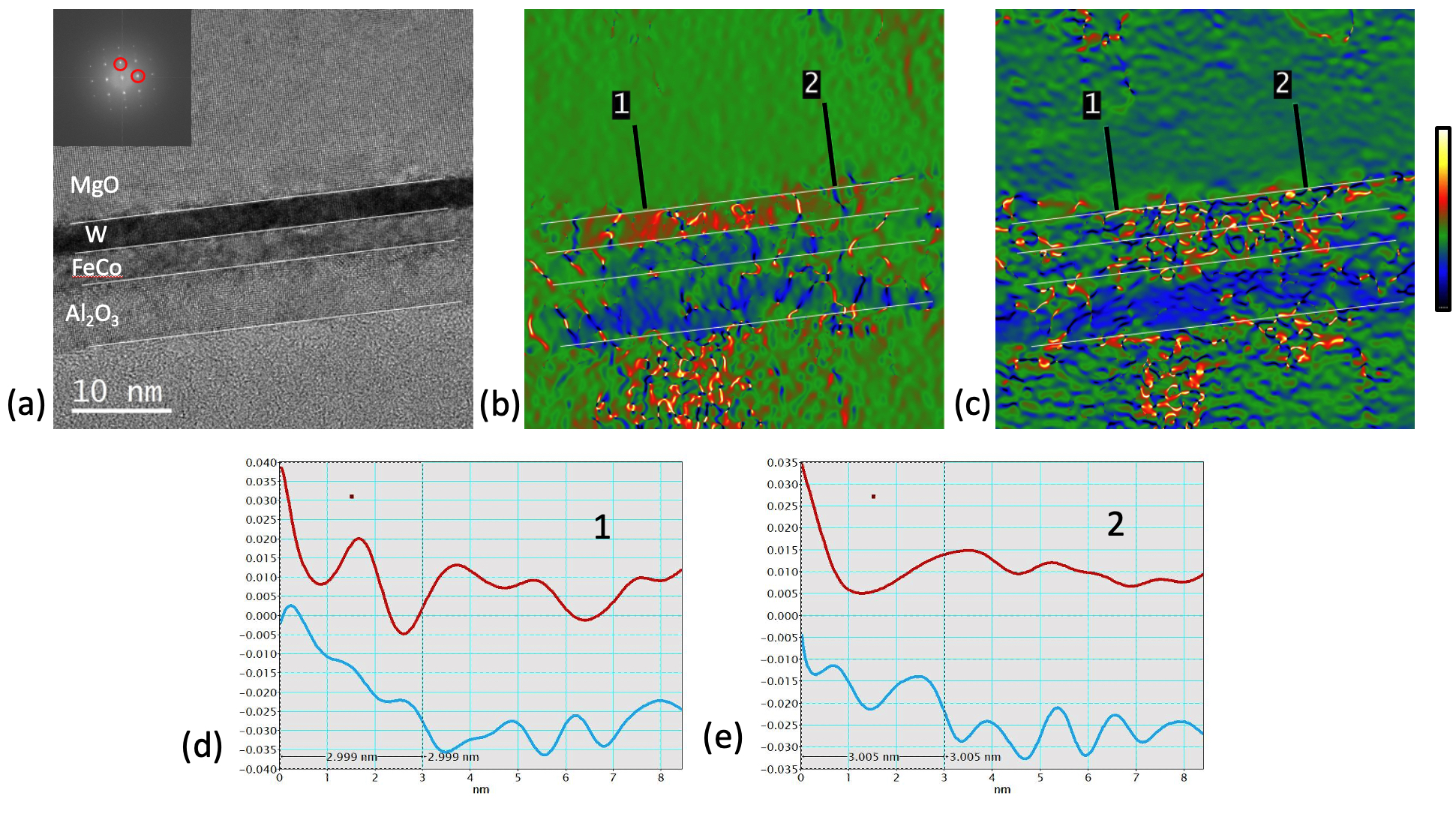}
    \caption{Quantitative strain analysis performed for the MgO/W/CoFe sample. (a) HRTEM image where the inset shows the FFT transform of the HRTEM image. The two red circles indicate the lattice spacing used in the geometric phase analysis. The strong variation of the colors in (b) and (c) represents strain maps along the interface and perpendicular to the interface, respectively. The color bar as legend indicate the strain from $-10\%$ to $10\%$. Two line profiles, selected at regions without observable dislocations in MgO, are shown in (d) and (e), where the red line corresponds to the strain along the interface and the blue line to the strain perpendicular to the interface. The strain measurement started from the MgO/$\alpha$-W interface, reaching an average strain level at about 3 nm away from the interface.}
    \label{fig:FigureS4}
\end{figure}
\clearpage

\begin{figure}[!ht]
        \centering
\includegraphics[width=12cm]{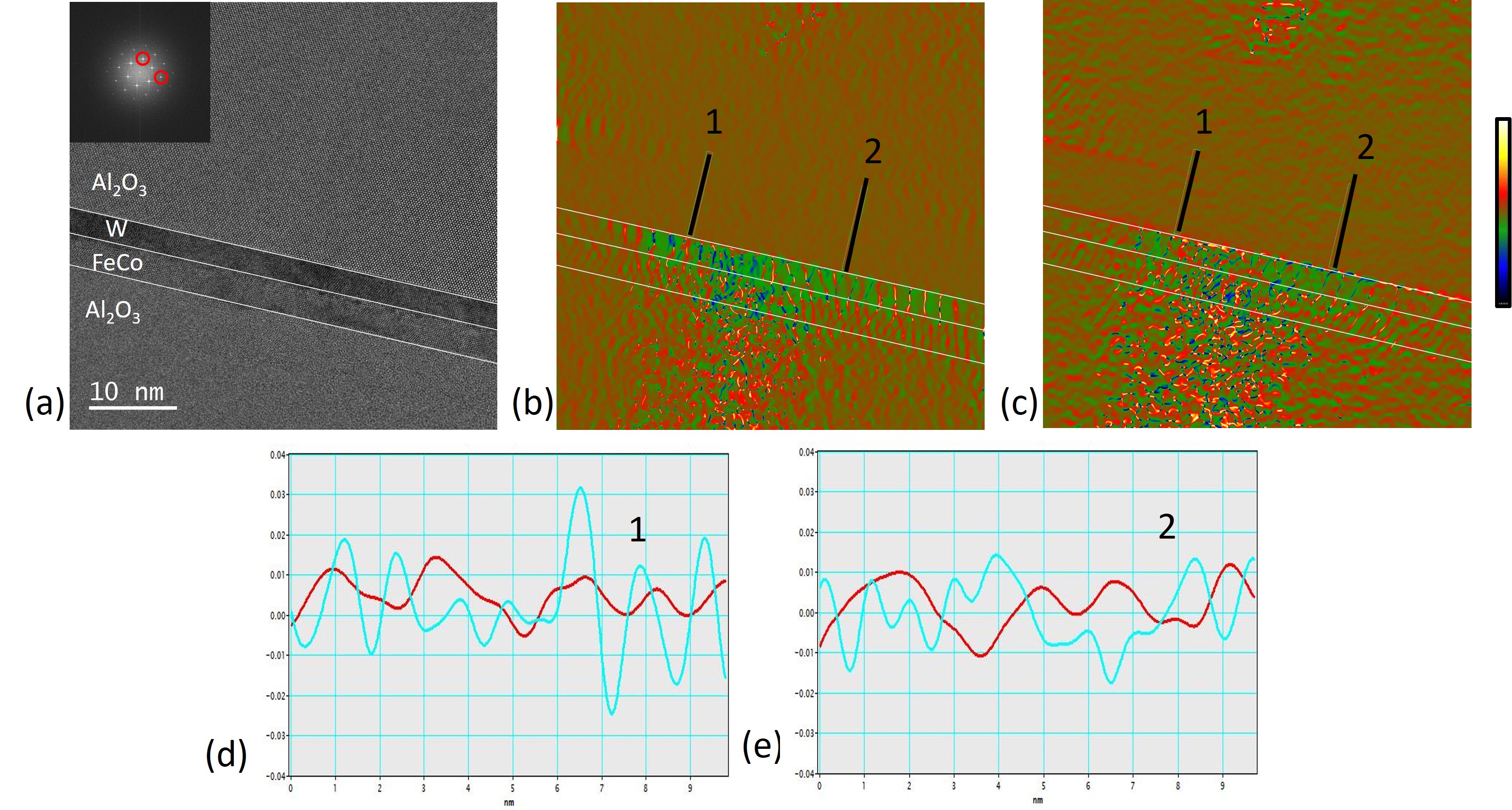}
    \caption{Quantitative strain analysis performed for the Al$_2$O$_3$ sample. (a) HRTEM of W/CoFe layers on Al$_2$O$_3$ substrate. Inset: FFT of the HRTEM image. The FFT components highlighted by red circle were used in the geometric strain analysis. (b) and (c) represents strain maps along the interface and perpendicular to the interface, respectively. The color bar as legend indicate the strain from $-10\%$ to $10\%$. Two line profiles, selected at regions without observable dislocations in Al$_2$O$_3$, are shown in (d) and (e), where the red line corresponds to the strain along the interface and the blue line to the strain perpendicular to the interface. The measurement started at the interface between Al$_2$O$_3$ and $\alpha$-W.}
    \label{fig:FigureS5}
\end{figure}
\clearpage

\section{Results from THz measurements}

\begin{figure}[!ht]
        \centering
    \includegraphics[width=10cm]{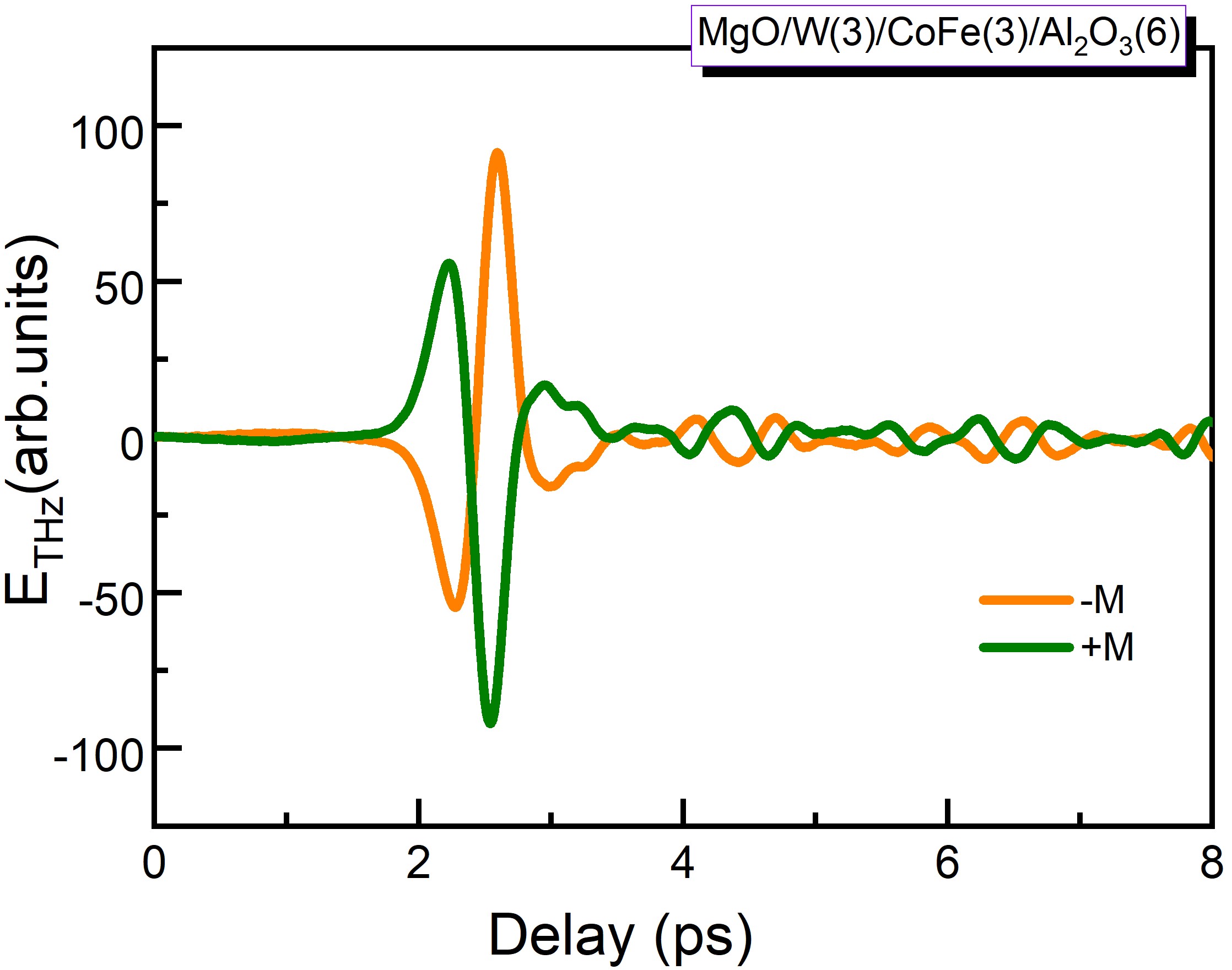}
    \caption{THz electric field in time domain for MgO/W($3$)/CoFe($3$) when pumping from substrate side using $0.3$ mJ$/$cm$^2$ laser fluency. The THz signal is inverted when the magnetization direction is reversed.}
    \label{fig:FigureS6}
\end{figure}

\begin{figure}[!ht]
        \centering
    \includegraphics[width=13cm]{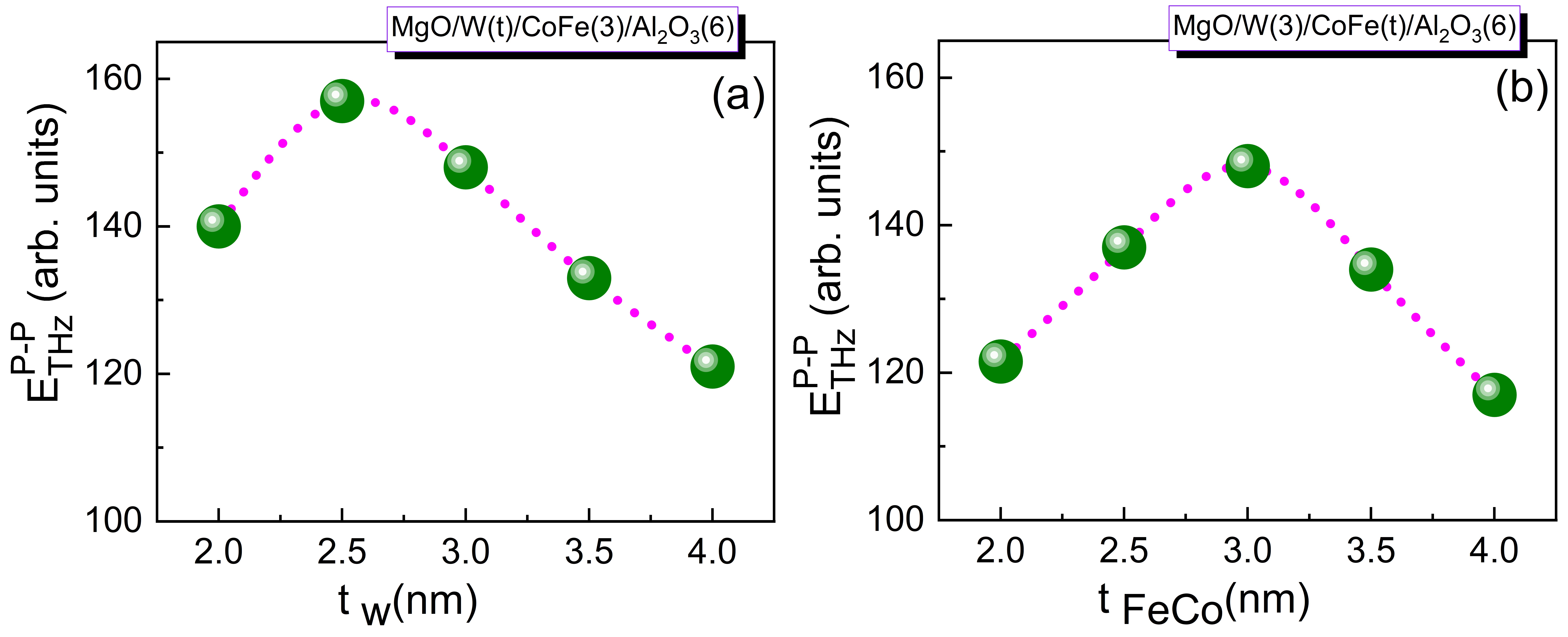}
    \caption{(a) THz electric field peak-to-peak amplitude for MgO/W($t_W$)/CoFe($3$) versus thickness $t_W$ of the tungsten layer. (b) THz electric field peak-to-peak amplitude for MgO/W($3$)/CoFe($t_{CoFe}$) versus thickness $t_{CoFe}$ of the CoFe layer.}
    \label{fig:FigureS7}
\end{figure}

\begin{figure}[!ht]
        \centering
    \includegraphics[width=12cm]{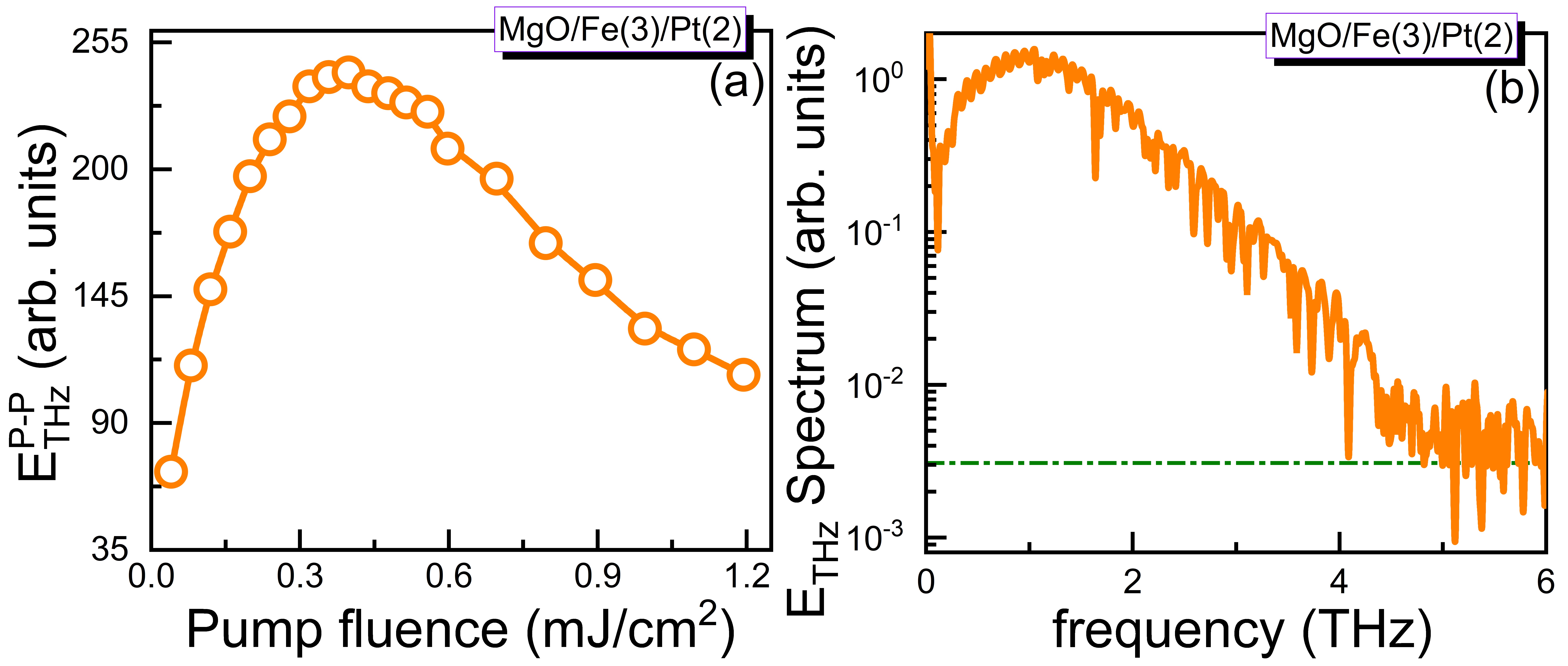}
    \caption{(a) THz electric field peak-to-peak amplitude for MgO/Fe(3)/Pt(2) when pumping from substrate side using $0.3$ mJ$/$cm$^2$ laser fluency. (b) Fast Fourier transform of time-domain signal for MgO/Fe(3)/Pt(2).}
    \label{fig:FigureS8}
\end{figure}

\newpage
\section{Transmittance, reflectance and absorption results}

\begin{figure}[!ht]
        \centering
    \includegraphics[width=12 cm]{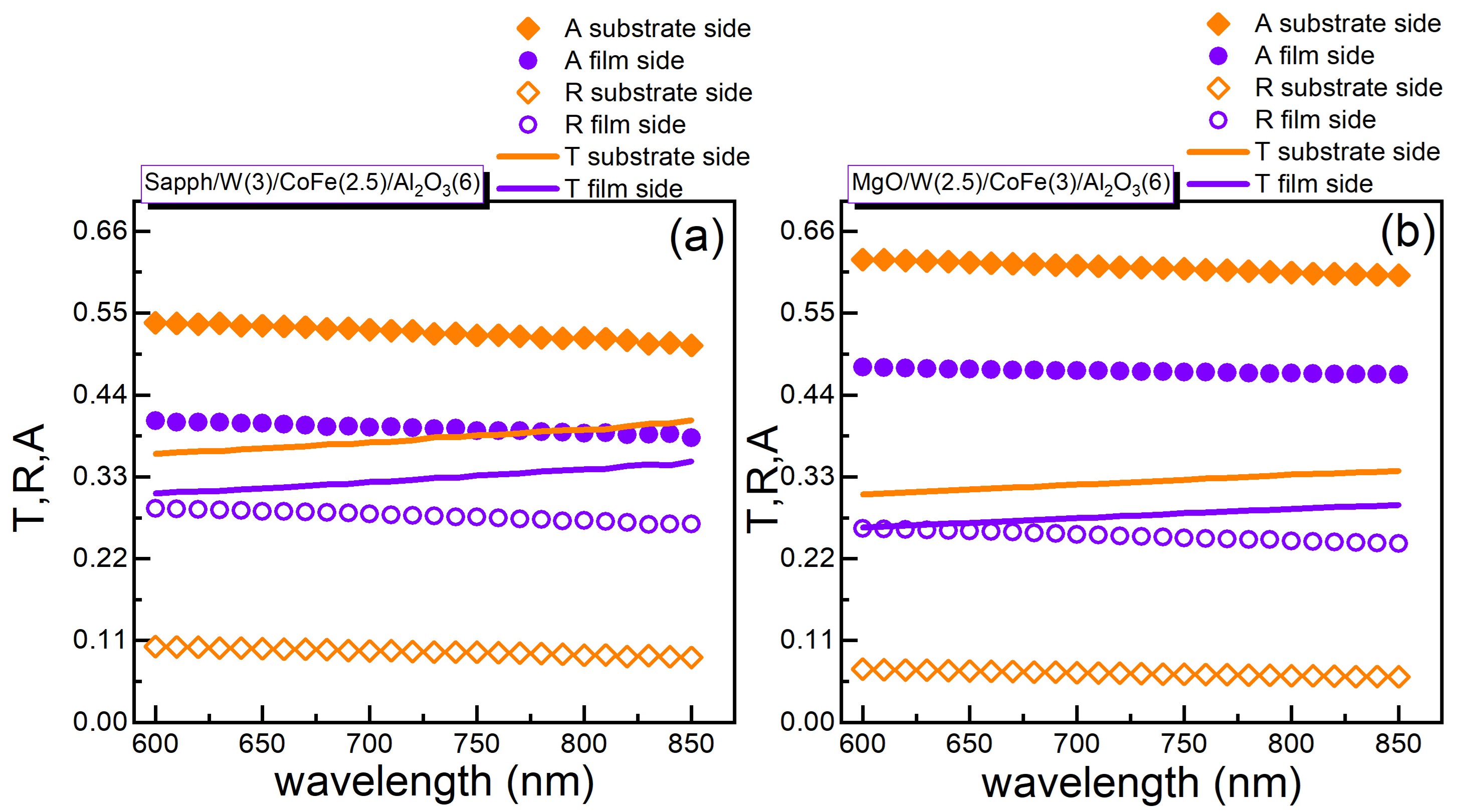}
    \caption{Transmittance ($T$), reflectance ($R$) and absorption ($A$) versus wavelength for (a) W($3$)/CoFe($2.5$)/Al$_2$O$_3$($6$) grown on single-side polished Al$_2$O$_3$ and (b) W($2.5$)/CoFe($3$)/Al$_2$O$_3$($6$) grown on single-side polished MgO.}
    \label{fig:FigureS9}
\end{figure}

\begin{figure}[!ht]
        \centering
    \includegraphics[width=12 cm]{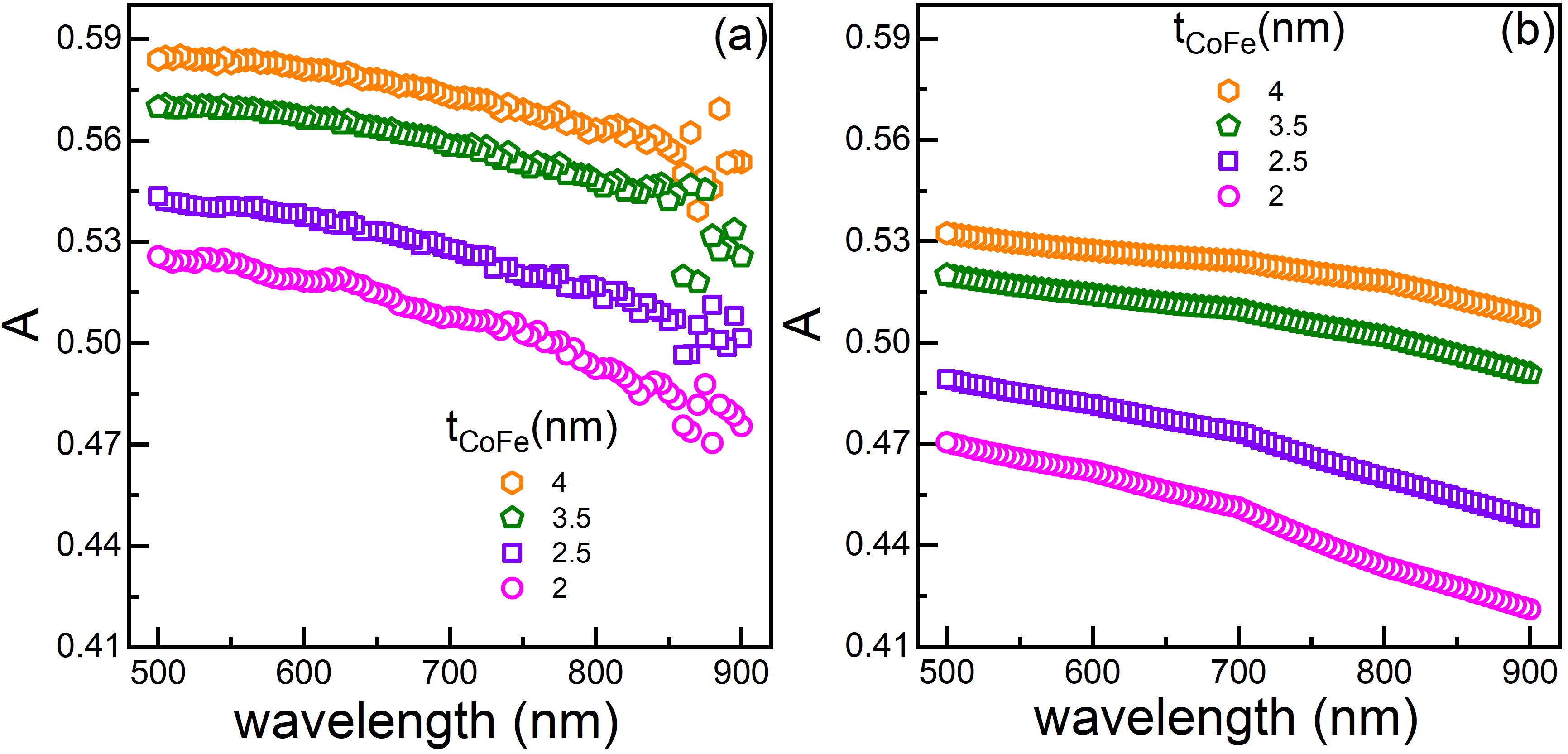}
    \caption{Absorption spectra for sapphire/W($3$)/CoFe($t_{CoFe}$)/Al$_2$O$_3$($6$); (a) experimental and (b) calculated results.}
    \label{fig:FigureS10}
\end{figure}

\clearpage


%

\section{Magnetization and resistivity results}
\begin{figure}[!ht]
    \centering
    \includegraphics[width=10cm]{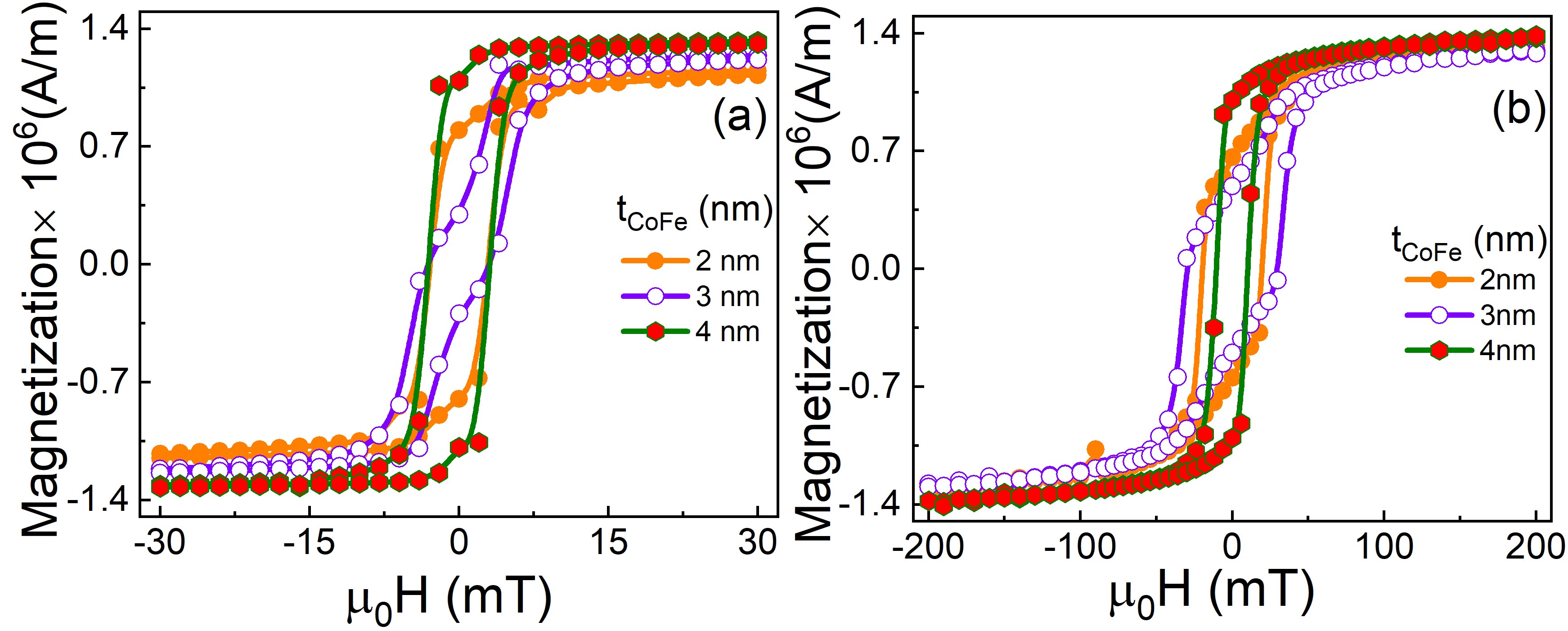}
    \caption{Thickness dependent magnetization of (a) MgO/W($3$)/CoFe($t_{CoFe}$) and (b) Al$_2$O$_3$/W($3$)/CoFe($t_{CoFe}$). The saturation magnetization values are $1.40 \times 10^6$ A/m, $1.34 \times 10^6$ A/m and $1.24 \times 10^6$ A/m for $t_{CoFe} = 4,3$ and $2$ nm, repectively for the MgO substrate; the saturation magnetization values are $1.43 \times 10^6$ A/m, $1.38 \times 10^6$ A/m and $1.38 \times 10^6$ A/m   for $t_{CoFe} = 4,3$ and $2$ nm, repectively for the sapphire substrate. The saturation magnetization values were calculated using the nominal thickness values.}
    \label{fig:Figure S11}
\end{figure}

\begin{figure}[!ht]
    \centering
    \includegraphics[width=5cm]{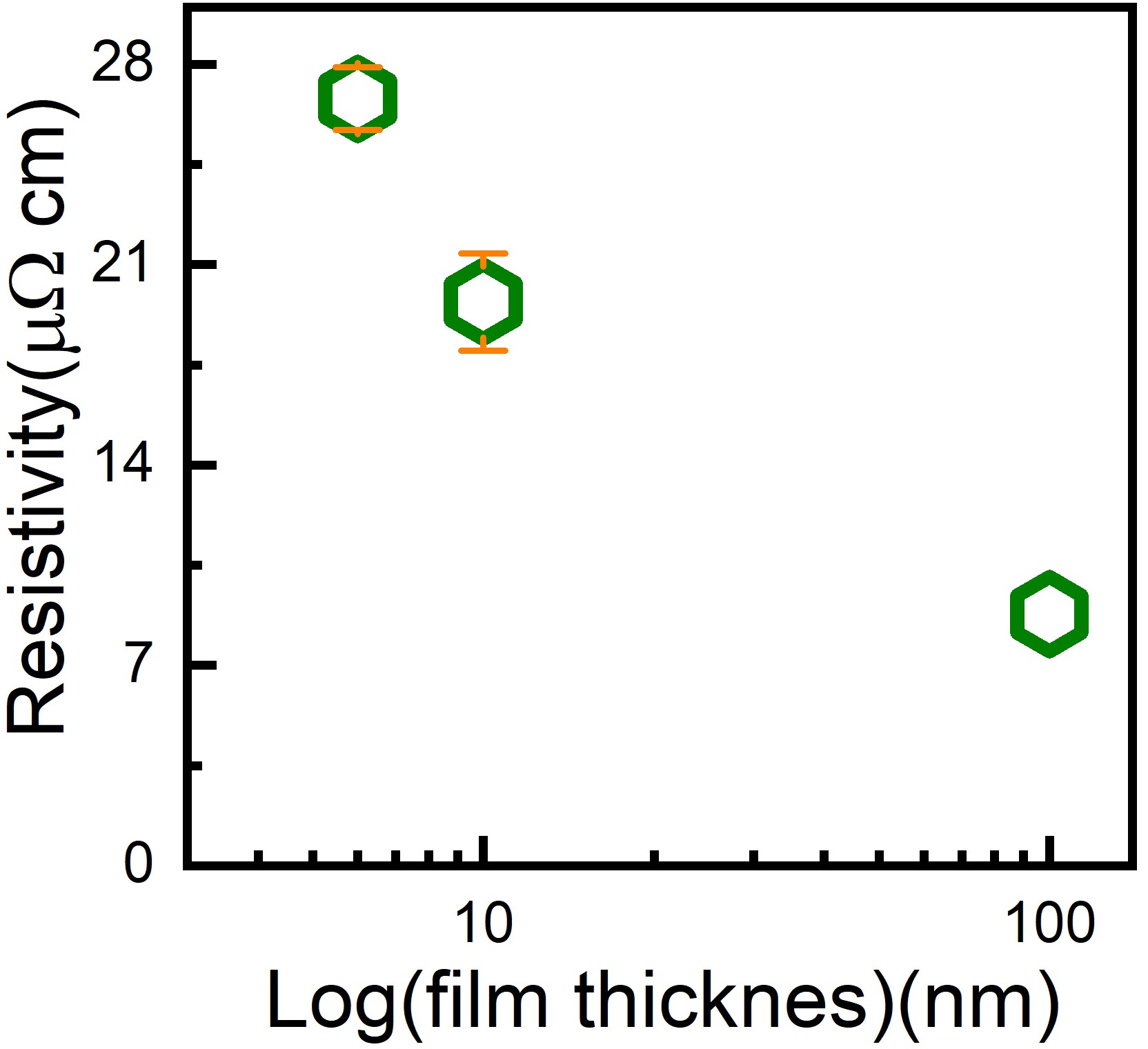}
    \caption{Thickness dependent electrical resistivity of $\alpha$-W. The increase at small thickness is due to the increased contribution from surface and interface electron scattering.}
    \label{fig:FigureS12}
\end{figure}

\clearpage

\section{Reflectance and transmittance experiments}

\begin{figure}[!ht]
    \centering
    \includegraphics[width=6cm]
    {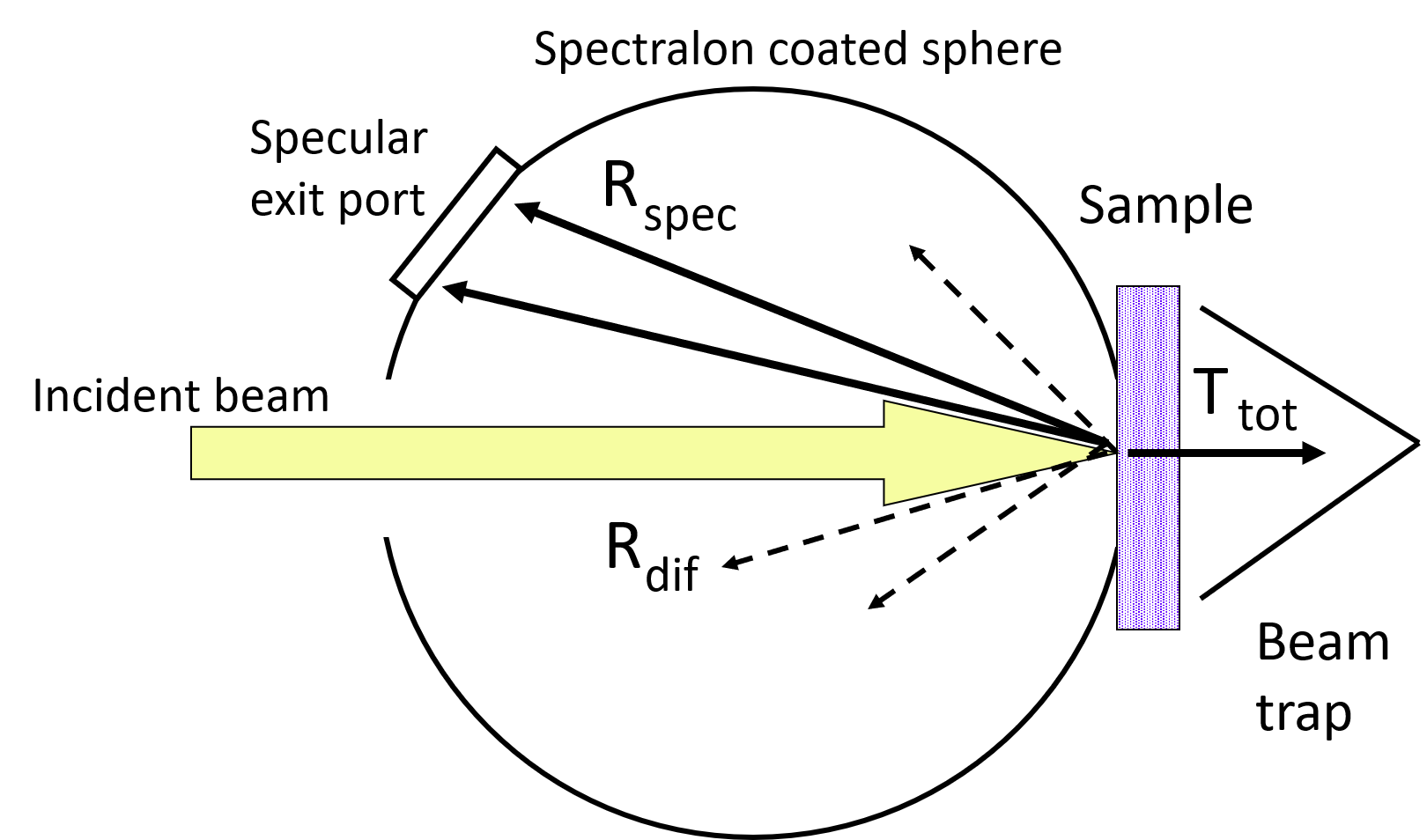}
    \includegraphics [width=6cm]
    {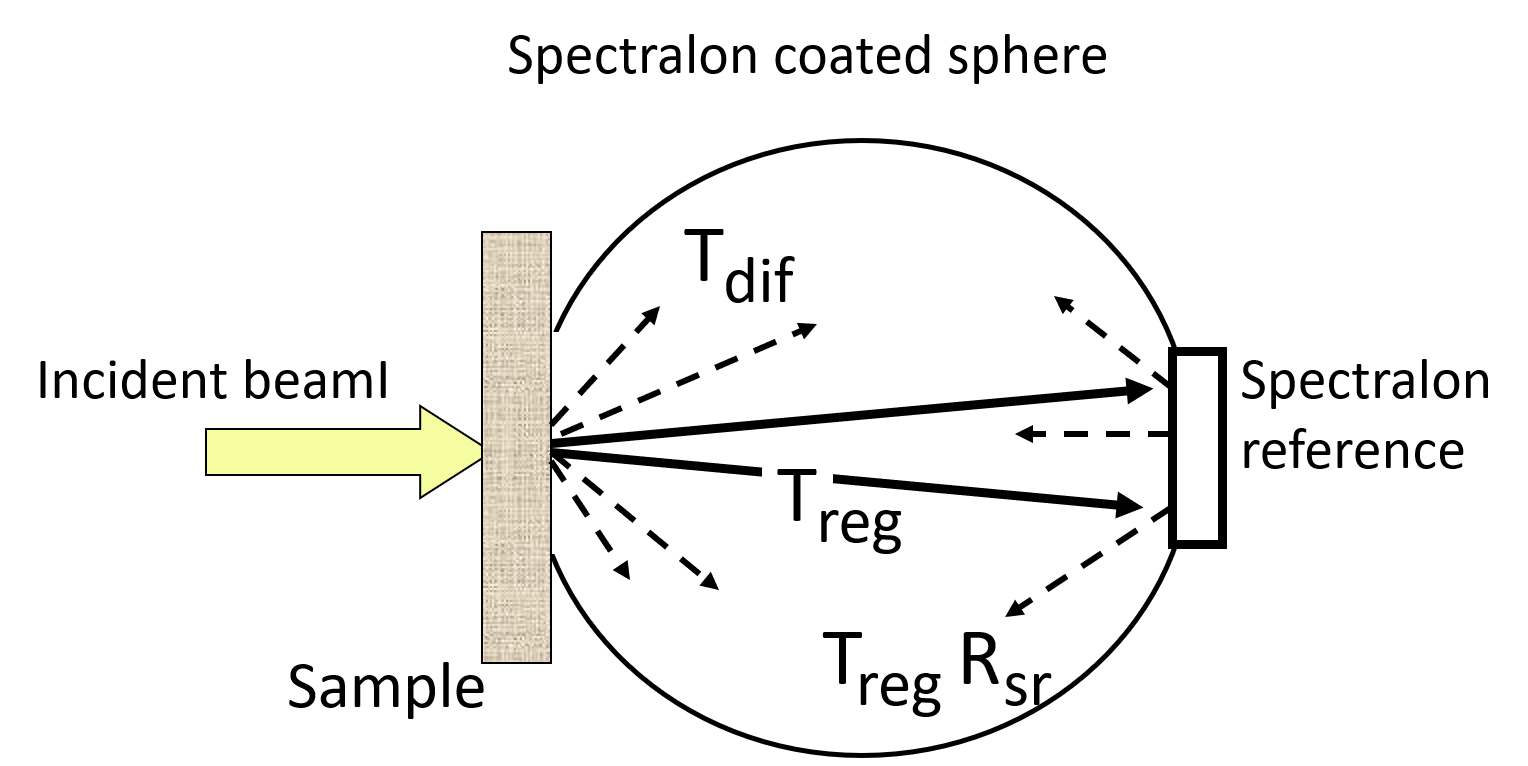}
    \caption{Left: Schematic illustration of the integrating sphere in reflectance mode. Right: Schematic illustration of the integrating sphere in transmittance mode.}
    \label{fig:FigureS13}
\end{figure}

Schematic illustrations of the sphere geometry in reflectance and transmittance modes are shown in Fig. \ref{fig:FigureS13}. 

\begin{figure}[!ht]
    \centering
    \includegraphics[width=12cm]{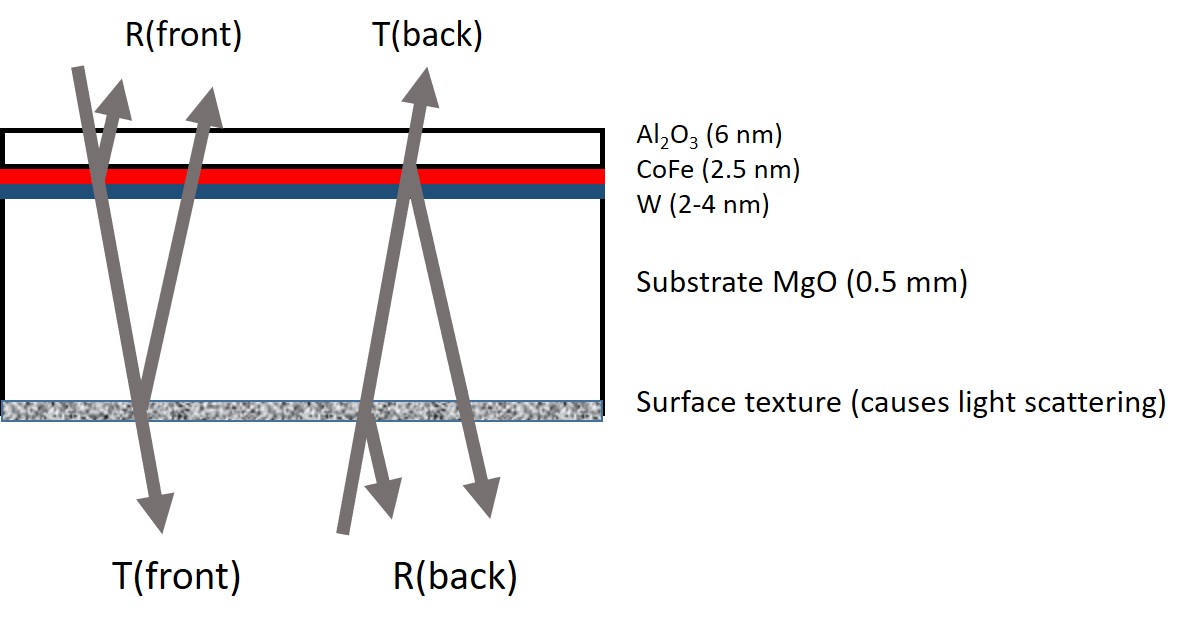}
        \caption{Schematic illustration of the film stacks used for the experiments.} \label {fig:FigureS14}
\end{figure}

The film stack configuration is illustrated in Fig. \ref{fig:FigureS14} with the measured reflected and transmitted beams shown. Subscript “film” is used when incident light hits the sample from the film side and subscript “substrate” is used when incident light hits the sample from the substrate side. There is a considerable difference in the two cases, as light beams reflected from the two surfaces of the substrate are non-coherent and are not subject to interference effects. The aluminium-oxide cap on the film side, on the other hand, is very thin and thus interference occurs between multiply reflected beams. The surface texture of the back surface of the substrate also causes light to be scattered in different ways in the front and back configurations. These two effects result in the fact that the transmittance and reflectance spectra are different for the two measurement orientations.

\clearpage

\section{X-ray reflectivity}

X-ray reflectivity (XRR) measurementsTo were used to obtain the thickness of W ($t_{W}^E$) and CoFe ($t_{CoFe}^E$), the substrate-film interface roughness ($\sigma_{substrate/W}$), the W-CoFe interface roughness ($\sigma_{W/CoFe}$) and the film-capping interface roughness ($\sigma_{CoFe/capping}$) in our emitters. The XRR spectra were recorded in the range of 2$\theta$=(0$^{\circ}-$6$^{\circ}$). The layer thickness and interface roughness were investigated by simulating the XRR spectra using the GenX software. The estimated values of the layer thicknesses and both interface roughnesses for all emitters are presented in Tables ST1-ST4. 
\begin{table}[!ht]
\small
  \caption{Comparison of nominal ($t_{W}^N$) and GenX extracted  thickness ($t_{W}^E$)  from XRR analysis for films grown on MgO(001) substrates. The nominal thickness of the CoFe layer was $t_{CoFe}^N = 30$ \AA. The thickness and roughness are in \AA.}  
  \begin{tabular*}{0.94\textwidth}{@{\extracolsep{\fill}}llllllll}
    \hline
    \hline
    & $t_{W}^N$ & $\sigma_{substrate/W}$ & $t_{W}^E$ & $\sigma_{W/CoFe}$ & $t_{CoFe}^E$ & $\sigma_{CoFe/Al_2O_3}$ & $t_{Al_2O_3}^E$\\
    \hline
    \hline
    & 40 & 3 & 38 & 1 & 26 & 5 & 58\\
    & 35 & 2 & 34 & 1 & 26 & 4 & 59\\
    & 30 & 2 & 29 & 1 & 29 & 12 & 56\\
    & 25 & 2 & 24 & 0 & 28 & 8 & 58\\
    & 20 & 3 & 20 & 1 & 26 & 5 & 59\\
    \hline
    \end{tabular*}
\end{table}

\begin{table}[!ht]
\small
  \caption{Comparison of nominal ($t_{W}^N$) and GenX extracted  thickness ($t_{W}^E$)  from XRR analysis for films grown on Al$_{2}$O$_{3}$($11\bar{2}0$) substrates. The nominal thickness of the CoFe layer was $t_{CoFe}^N = 30$ \AA. The thickness and roughness are in \AA.}  \label{tbl:XRR and thickness}
  \begin{tabular*}{0.94\textwidth}{@{\extracolsep{\fill}}llllllll}
    \hline
    \hline
    & $t_{W}^N$ & $\sigma_{substrate/W}$ & $t_{W}^E$ & $\sigma_{W/CoFe}$ & $t_{CoFe}^E$ & $\sigma_{CoFe/Al_2O_3}$ & $t_{Al_2O_3}^E$\\
    \hline
    \hline
    & 40 & 5 & 41 & 5 & 25 & 5 & 58\\
    & 35 & 1 & 35 & 4 & 25 & 4 & 57\\
    & 30 & 8 & 31 & 5 & 26 & 6 & 57\\
    & 25 & 1 & 25 & 3 & 26 & 5 & 57\\
    & 20 & 10 & 21 & 3 & 24 & 4 & 60\\
    \hline
    \end{tabular*}
\end{table}

\begin{table}[!ht]
\small
  \caption{Comparison of nominal ($t_{CoFe}^N$) and GenX extracted  thickness ($t_{CoFe}^E$)  from XRR analysis for films grown on MgO(001) substrates. The nominal thickness of the W layer was $t_{W}^N = 30$ \AA. The thickness and roughness are in \AA.}  
  \begin{tabular*}{0.94\textwidth}{@{\extracolsep{\fill}}llllllll}
    \hline
    \hline
    & $t_{CoFe}^N$ & $\sigma_{substrate/W}$ & $t_{W}^E$ & $\sigma_{W/CoFe}$ & $t_{CoFe}^E$ & $\sigma_{CoFe/Al_2O_3}$ & $t_{Al_2O_3}^E$\\
    \hline
    \hline
    & 40 & 1 & 29 & 1 & 30 & 6 & 64\\
    & 35 & 2 & 29 & 1 & 29 & 8 & 61\\
    & 30 & 2 & 29 & 1 & 29 & 12 & 56\\
    & 25 & 2 & 29 & 1 & 19 & 2 & 63\\
    & 20 & 3 & 28 & 2 & 18 & 3 & 65\\
    \hline
    \end{tabular*}
\end{table}

\begin{table}[!ht]
\small
  \caption{Comparison of nominal ($t_{CoFe}^N$) and GenX extracted  thickness ($t_{CoFe}^E$)  from XRR analysis for films grown on Al$_{2}$O$_{3}$($11\bar{2}0$) substrates. The nominal thickness of the W layer was $t_{W}^N = 30$ \AA. The thickness and roughness are in \AA.}  
  \begin{tabular*}{0.94\textwidth}{@{\extracolsep{\fill}}llllllll}
    \hline
    \hline
    & $t_{CoFe}^N$ & $\sigma_{substrate/W}$ & $t_{W}^E$ & $\sigma_{W/CoFe}$ & $t_{CoFe}^E$ & $\sigma_{CoFe/Al_2O_3}$ & $t_{Al_2O_3}^E$\\
    \hline
    \hline
    & 40 & 2 & 31 & 3 & 33 & 5 & 58\\
    & 35 & 10 & 30 & 4 & 28 & 4 & 58\\
    & 30 & 8 & 31 & 5 & 26 & 6 & 57\\
    & 25 & 10 & 30 & 3 & 20 & 4 & 59\\
    & 20 & 10 & 30 & 3 & 19 & 6 & 54\\
    \hline
    \end{tabular*}
\end{table}

\begin{sidewaystable*}[!ht]  
\caption{Influence of structural order on the THz emission amplitude in various material systems [Epi: Epitaxial; P-Epi: Partial Epitaxial; Am: Amorphous; Poly: Polycrystalline; AFM: Antiferromagnetic; FM: Ferromagnetic; HM: Heavy Metal]}


\begin{tabular}{
  >{\centering\arraybackslash}p{2.5cm}   
  >{\centering\arraybackslash}p{3.3cm}   
  >{\centering\arraybackslash}p{3.7cm}   
  >{\centering\arraybackslash}p{7.7cm}   
  c                                      
}

\toprule
\textbf{MATERIAL SYSTEM} & 
\textbf{STRUCTURAL STATE} & 
\textbf{COMPARISON METRIC} & 
\textbf{RESULT} & 
\textbf{REF.} \\
\midrule

Fe/Pt on MgO or Al$_2$O$_3$ & 
 Epi Fe/MgO, Poly Fe/Al$_2$O$_3$ & 
Effect of Fe layer Epi & 
$E_{\mathrm{THz}}$(Epi Fe) $>$ $E_{\mathrm{THz}}$(Poly Fe) & \cite{torosyan2018optimized,nenno2019modification} \\

Fe/Pt & 
 Epi-Fe/Epi or P-Epi Pt & 
 Pt layer Epi quality & 
$E_{\mathrm{THz}}$(Epi Pt) = 2 $\times$ $E_{\mathrm{THz}}$(P-Epi Pt) & 
\cite{nenno2019modification}  \\

Ta/CoFeB/MgO & 
Am, Poly CoFeB  & 
Am vs Poly CoFeB nature & 
$E_{\mathrm{THz}}$(Poly CoFeB) = 1.5$\times$ $E_{\mathrm{THz}}$(Am CoFeB) & 
\cite{sasaki2017annealing}  \\

Co/Pt & 
Epi, Poly Co & 
Epi vs Poly Co nature & 
$E_{\mathrm{THz}}$(Epi Co) $>$ $E_{\mathrm{THz}}$(Poly Co) & 
\cite{PhysRevMaterials.3.084415} \\

Fe or CoFeB /Pt & 
Epi Fe, Am CoFeB  & 
Fe vs CoFeB FM nature & 
$E_{\mathrm{THz}}$(Am CoFeB/Pt) $\approx$ $E_{\mathrm{THz}}$(Epi Fe/Pt) & 
\cite{seifert2016efficient} \\

Fe or CoFeB/$\alpha$-Ta & 
Epi Fe, Poly CoFeB & 
Fe vs CoFeB FM nature & 
$E_{\mathrm{THz}}$(Epi Fe/$\alpha$-Ta) = 3$\times$ $E_{\mathrm{THz}}$(Poly CoFeB/$\alpha$-Ta) & 
\cite{kumar2021optical} \\

Pt or W/Co & 
  State is not specified  & 
Pt vs W (HM comparison) & 
$E_{\mathrm{THz}}$(Pt/Co) = 2$\times$ $E_{\mathrm{THz}}$(W/Co) & 
\cite{Wu2017} \\

Pt or W/NiO & 
 Epi NiO, HM state is not specified & 
Pt vs W (HM comparison) & 
$E_{\mathrm{THz}}$(Pt/NiO) = 3$\times$ $E_{\mathrm{THz}}$(W/NiO) & 
\cite{qiu2021ultrafast} \\

Annealed Fe/Pt & 
All layers are Epi & 
As-grown vs annealed Fe/Pt (L1$_0$-FePt phase interface) & 
Enhanced $E_{\mathrm{THz}}$ with L1$_0$ phase & 
\cite{scheuer2022thz} \\

Fe/Pt/MgO or MAO & 
All layers are Epi  & 
Substrate-induced strain in Epi stack & 
$E_{\mathrm{THz}}$(Fe/Pt/MgO) $>$ $E_{\mathrm{THz}}$(Fe/Pt/MAO) & 
\cite{gupta2021strain} \\

NiO/Pt/MgO or Al$_2$O$_3$ & 
All layers are Epi & 
Substrate-induced structural variation & 
$E_{\mathrm{THz}}$(NiO/MgO) = 10$\times$ $E_{\mathrm{THz}}$(NiO/Al$_2$O$_3$) & 
\cite{agarwal2023strong} \\

Mn$_3$Sn(0001 or 11$\bar{2}$0)/Co or Pt & 
Co,Pt state is not specified & 
AFM crystallographic orientation dependence & 
$E_{\mathrm{THz}}$(Mn$_3$Sn(11$\bar{2}$0)/Co) $>$ $E_{\mathrm{THz}}$(Mn$_3$Sn(0001)/Co);
$E_{\mathrm{THz}}$(Mn$_3$Sn(11$\bar{2}$0)/Pt) $<$ $E_{\mathrm{THz}}$(Mn$_3$Sn(0001)/Pt) & 
\cite{zhou2019orientation} \\

$\alpha$-Fe$_2$O$_3$/Pt/Al$_2$O$_3$ & 
All layers are Epi & 
AFM crystallographic orientation dependence  & 
$E_{\mathrm{THz}}$($\alpha$-Fe$_2$O$_3$(1$\bar{1}$02)/Pt) $>$ \(E_{\mathrm{THz}}\) ($\alpha$-Fe$_2$O$_3$(0001 or 11$\bar{2}$0)/Pt) & 
\cite{liu2021strain}\\

$\alpha$-W/CoFe on MgO or Al$_2$O$_3$ \& Fe/Pt & 
All layers are Epi & 
Structural differences within Epi series & 
$E_{\mathrm{THz}}$($\alpha$-W(110)/CoFe(110)) $>$ $E_{\mathrm{THz}}$($\alpha$-W(001)/CoFe(001)); 
$E_{\mathrm{THz}}$($\alpha$-W/CoFe) $\approx$ $E_{\mathrm{THz}}$(Fe/Pt) & 
Our result \\

\botrule
\end{tabular}
\end{sidewaystable*}

\clearpage
\bibliographystyle{unsrt}
\bibliography{sn-bibliography}